\documentclass[12pt]{article}
\usepackage{epsfig, epsf, graphicx, subfigure}
\usepackage{pstricks, pst-node, psfrag}
\usepackage{amssymb,amsmath,bm}
\usepackage{verbatim,enumerate}
\usepackage{rotating, lscape}
\usepackage{setspace}
\usepackage[english]{babel}
\usepackage[sort&compress]{natbib}
\usepackage{multirow}
\usepackage{theorem}

\def\d{\mathrm{d}}

\def\p{\partial}

\def\00{\mathrm{0}}
\def\ZZ{\mathbf{Z}}

\def\AA{\mathcal{A}}
\def\YY{\boldsymbol{Y}}

\def\AB{\boldsymbol{A}}

\def\hh{\boldsymbol{h}}
\def\tht{\boldsymbol{\theta}}

\def\uu{\mathbf{u}}

\def\xx{\mathbf{x}}

\def\ss{\boldsymbol{s}}

\newtheorem{prop}{Proposition}
\newtheorem{corol}{Corollary}

\begin{document}
	
\thispagestyle{empty} \baselineskip=28pt \vskip 5mm
\begin{center} 
\Huge{\bf Max-convolution processes with random shape indicator kernels}
\end{center}
	
\baselineskip=12pt \vskip 10mm
	
\begin{center}\large
		Pavel Krupskii\footnote[1]{\baselineskip=10pt University of Melbourne, Parkville, Victoria, 3010, Australia. E-mail: pavel.krupskiy@unimelb.edu.au.}, 
		Rapha\"{e}l Huser\footnote[2]{\baselineskip=10pt Computer, Electrical and Mathematical Sciences and Engineering (CEMSE) Division, King Abdullah University of Science and Technology (KAUST), Thuwal 23955-6900, Saudi Arabia. E-mail: raphael.huser@kaust.edu.sa.}
\end{center}
	
\baselineskip=17pt \vskip 10mm \centerline{\today} \vskip 15mm
	
\begin{center}
	{\large{\bf Abstract}}
\end{center}
	
In this paper, we introduce a new class of models for spatial data obtained from max-convolution processes based on indicator kernels with random shape. We show that this class of models have appealing dependence properties including tail dependence at short distances and independence at long distances. We further consider max-convolutions between such processes and processes with tail independence, in order to separately control the bulk and tail dependence behaviors, and to increase flexibility of the model at longer distances, in particular, to capture intermediate tail dependence. 
We show how parameters can be estimated using a weighted pairwise likelihood approach, and we conduct an extensive simulation study to show  that the proposed inference approach is feasible in high dimensions and it yields accurate parameter estimates in most cases. We apply the proposed methodology to analyse daily temperature maxima measured at 100 monitoring stations in the state of Oklahoma, US. Our results indicate that our proposed model provides a good fit to the data, and that it captures both the bulk and the tail dependence structures accurately.

	
\baselineskip=14pt

\par\vfill\noindent
	
\clearpage\pagebreak\newpage \pagenumbering{arabic}
\baselineskip=17pt


\section{Introduction}
\label{intro}

The statistical modeling of natural hazards and spatial extreme events requires specialized models that appropriately capture the joint tail behavior \citep{Huser.Davison2014, Huser.Wadsworth2017,Xu.Genton2016, Genton.Padoan.ea2015}. As assessment of future risks relies on tail extrapolation, it is indeed crucial to develop models that are robust for reliable tail extrapolation, while at the same time being flexible enough to adapt to the various asymptotic regimes that the data may exhibit \citep{Huser.Wadsworth:2020}. 


From this perspective, spatial processes can be classified into two broad model classes: processes exhibiting tail-dependence and those exhibiting tail-independence (in an asymptotic sense). Loosely speaking, the former allow for the most extreme events (in the limit) to occur simultaneously at different locations, while the latter do not. In practical terms, extremes from tail-dependent processes can have quite a large spatial extent and thus large-scale impacts, while they tend to be much more localized under tail-independence, especially as the magnitude of the extreme event intensifies. Mathematically, to characterize the tail dependence class, it is helpful to consider processes on a standardized scale, e.g., the uniform ${\rm Unif}(0,1)$ scale. Assume that the spatial process of interest, $Z(\ss)$, $\ss\in\mathcal{S}$, is stationary with continuous marginal distributions $F_Z(z)$, and consider the standardized process $U(\ss)=F_Z\{Z(\ss)\}$ with common standard uniform margins, thus focusing on the \emph{copula} structure of $Z(\ss)$. If, for any pair of sites $\{\ss_1,\ss_2\}\subset\mathcal S$, the limiting probability $\lambda_U(\boldsymbol{h})=\lim_{u\to1}\Pr\{U(\ss_1)>u\mid U(\ss_2)>u\}$, $\boldsymbol{h}=\boldsymbol{s}_1-\boldsymbol{s}_2$, exists and is positive, i.e., $\lambda_U(\boldsymbol{h})>0$, then both $Z(\ss)$ and $U(\ss)$ are called (asymptotically) tail-dependent. By contrast, if this limit probability equals zero, i.e., $\lambda_U(\boldsymbol{h})=0$, they are called (asymptotically) tail-independent. While $\lambda_U(\boldsymbol{h})$ characterizes the limiting form of dependence in the upper tail, we can also define by symmetry a similar coefficient, $\lambda_L(\boldsymbol{h})=\lim_{u\to0}\Pr\{U(\ss_1)<u\mid U(\ss_2)<u\}$ for the lower tail, and an important methodological problem is to develop spatial models that have flexible forms of dependence in both tails, i.e., potentially asymmetric \citep{Gong.Huser:2022} and/or changing as a function of distance $\boldsymbol{h}$ between sites \citep{Wadsworth.Tawn:2012b}.

Classical geostatistical models and popular extreme-value models are usually quite limited in their ability to capture joint tail characteristics, and even the most recently proposed models often have strong restrictions \citep{Huser.Wadsworth:2020}. Some models are always asymptotically tail-independent, such as the wide class of trans-Gaussian processes \citep{Xu.Genton2017},  inverted max-stable processes \citep{Wadsworth.Tawn:2012b} and Laplace random processes \citep{Opitz2015}, while others are always asymptotically tail-dependent, such as certain types of non-Gaussian latent factor processes \citep{Krupskii.Huser.ea2016}, max-stable processes \citep{Schlather.2002, Huser.Genton2016} and Pareto processes \citep{Fereira.DeHaan2014, deFondeville.Davison.2018} popularly used for modeling spatial extremes. Other more recent models provide improvements in their ability to capture both asymptotic tail-dependence and tail-independence in a rather flexible way, but they cannot capture full independence as the distance between sites increases arbitrarily \citep{Huser.Opitz.ea2016,Huser.Wadsworth2017}. Other models can capture a change of asymptotic tail-dependence class as a function of the distance between sites, as well as full independence at infinity, such as max-mixture models \citep{Wadsworth.Tawn:2012b}, or the spatial conditional extremes model \citep{Wadsworth.Tawn2022}, or the so-called SHOT model of \citet{Hazra.etal:2021}, but  they also have other limitations; in particular, existing max-mixture models are often relatively heavily parameterized, which complicates inference; the spatial conditional extremes model does not possess a convenient ``unconditional representation''; and the SHOT model has intrinsic non-stationary artefacts. There is thus a need to develop stationary spatial models that possess high tail flexibility, that can capture full independence as the distance increases to infinity, while at the same time, allowing for fast inference and simulation.

Recently, Cauchy kernel convolution processes \citep{Krupskii.Huser2022} have been proposed to address some of these challenges, including the ability to capture a change in asymptotic tail-dependence class as a function of distance when considering compactly-supported kernel functions. However, these models and the spatial process mixture extension proposed in \cite{Krupskii.Huser2022} are still quite restrictive in the sub-asymptotic tail-independence structure that they can capture at large distances. Precisely, in the tail-independence case, the proposed model has a fast joint tail decay rate that is equivalent to that of white noise.

In this paper, we address these shortcomings by building upon both kernel convolutions \citep{Krupskii.Huser2022} and max-mixture constructions \citep{Wadsworth.Tawn:2012b}, in order to design new relatively parsimonious spatial models with a highly flexible tail structure, and which lead to amenable pairwise likelihood-based inference. Specifically, we here consider max-convolutions of the form
\begin{equation}
\label{eq2}
Z(\ss) = h(|\AA(\ss)|)\sup_{\ss^* \in \AA(\ss)} W(\d\ss^*), \quad \ss \in \mathbb{R}^q,
\end{equation} 
where $\AA(\ss) \subset \mathbb{R}^q$ is a compact subset of $\mathbb{R}^q$ with random shape such that $|\AA(\ss)| \leq A < \infty$ almost surely for any $\ss \in \mathbb{R}^q$, $W$ is a L\'evy process \citep{Sato1999} with independent increments, and $h(\cdot)$ is a continuous function on $[0, A]$. Motivated by the copula literature, we shall study the process \eqref{eq2} on a standardized scale in order to extract its dependence structure and disregard its marginal distributions. As we shall show, the proposed process \eqref{eq2} has attractive properties, namely: (i) it possesses tail-dependence at short distances and tail-independence at long distances; and (ii) the range of tail-dependence and the dependence decay rate can be separately controlled using parameters of the process $\AA(\ss)$.  We then further consider max-convolutions of the process \eqref{eq2} with a different, tail-independent process, in order to increase flexibility at longer distances and in particular, to capture intermediate dependence (i.e., a form of tail-independence, made precise below, that is weaker than tail-dependence but stronger than full independence). Furthermore, inference for all the proposed models can be performed relatively easily using a weighted pairwise likelihood approach. 


The rest of the paper is organized as follows. Section 2 presents our proposed modeling framework, with Section~\ref{subsec1.0} defining our model precisely, Section~\ref{subsec1.1} detailing the tail properties of the process \eqref{eq2}, Section~\ref{subsec1.2} focusing on a special case which retains flexibility in the tails and makes inference easier, and  Section~\ref{subsec1.3} considering max-mixture extensions that allow for greater flexibility at sub-asymptotic levels.  We discuss inference methods for these processes in Section~\ref{sec-infer}, and assess the performance of the proposed estimators by simulation in Section~\ref{sec-sim}. In Section 5, we apply the proposed models to analyze wind speed data, and Section 6 concludes with some discussion about future research directions. 

\section{Max-convolution processes}
\label{sec1}

\subsection{Model definition}
\label{subsec1.0}

Consider the process $Z(\ss)$ defined as in \eqref{eq2}. To construct the process on a standardized scale, we assume, without loss if generality, that the marginal distribution of $W$ is  Fr\'echet such that $\sup_{\ss^* \in \AA(\ss)} W(\d s^*)$ has the standard Fr\'echet distribution with the cumulative distribution function (cdf) $\Pr(\sup_{\boldsymbol{s}^*\in\mathcal{A}(\boldsymbol{s})}W({\rm d}\boldsymbol{s}^*) \leq z) = \exp(-1/z)$, $z > 0$, if $|\AA(\ss)| = 1$. 

Consider the random vector $(Z_{1}, Z_{2})^{\top} = (Z(\ss_1), Z(\ss_2))^{\top}$. 
For $\ss_1, \ss_2 \in \mathbb{R}^q$, let $A_i = |\AA_i|$, $i=1,2$, and $A_{12} = |\AA_{12}|$ denote random variables measuring the area of the disjoint random sets $\AA_1 = \AA(\ss_1) \backslash \AA(\ss_2)$, $\AA_2 = \AA(\ss_2) \backslash \AA(\ss_1)$, and $\AA_{12} = \AA(\ss_1) \cap \AA(\ss_2)$, respectively. We assume that the joint probability density function (pdf) of the random vector $\AB = (A_1, A_2, A_{12})^{\top}$ exists and we denote it by $g(\boldsymbol{x})$, $\boldsymbol{x}\in (0,\infty)^3$. Conditional on $\AB = \xx = (x_1, x_2, x_{12})^{\top}$, we can thus write, for $i=1,2$, 
$$Z_i = h_i(\xx) \max\{W_i(\xx), W_{12}(\xx)\}, \quad h_i(\xx) = h(x_i + x_{12}),$$
where $W_1(\xx), W_2(\xx)$ and $W_{12}(\xx)$ are Fr\'echet random variables distributed as 
$$
\Pr\{W_i(\xx) \leq z\} = e^{-x_i/z}, \quad \Pr\{W_{12}(\xx) \leq z\} = e^{-x_{12}/z}, \qquad z>0. 
$$
The joint cdf of the random vector $(Z_{1}, Z_{2})^{\top}$ can thus be expressed as
\begin{eqnarray}
\label{eq-bivcdf}
F_{12}(z_1, z_2) &=& \int_{(0,\infty)^3}\Pr\left[h_1(\xx) \max\{W_1(\xx), W_{12}(\xx)\} \leq z_1, h_2(\xx) \max\{W_2(\xx), W_{12}(\xx)\} \leq z_2\right]g(\xx)\d\xx \nonumber\\
&=& \int_{(0,\infty)^3} \Pr\left[W_1(\xx) \leq \frac{z_1}{h_1(\xx)}, W_2(\xx) \leq \frac{z_2}{h_2(\xx)}, W_{12}(\xx) \leq \min\left\{\frac{z_1}{h_1(\xx)}, \frac{z_2}{h_2(\xx)}\right\}\right] g(\xx)\d\xx \nonumber\\
&=& \int_{(0,\infty)^3} \exp\left[-\frac{x_1h_1(\xx)}{z_1}-\frac{x_2h_2(\xx)}{z_2} - x_{12}\max\left\{\frac{h_1(\xx)}{z_1}, \frac{h_2(\xx)}{z_2}\right\} \right]g(\xx)\d\xx,
\end{eqnarray}
and its marginal cdf is
\begin{equation}
\label{eq-univcdf}
F_i(z) = \Pr(Z_i \leq z) =  \int_{(0,\infty)^3} \exp\left\{-\frac{(x_i + x_{12})h_i(\xx)}{z}\right\}g(\xx)\d\xx, \quad i = 1,2.
\end{equation}
To estimate model parameters using the pairwise likelihood approach, one needs to compute the bivariate and marginal cdfs as given in (\ref{eq-bivcdf}) and (\ref{eq-univcdf}). While these functions are not available in closed form in most cases, numerical integration methods can be used to approximate them accurately in practice. In Section~\ref{subsec1.2}, we consider a simpler special case where inference can be made more easily.
 
\subsection{Tail properties}
\label{subsec1.1}

Following the notation introduced in the previous section, let $\bm Z=(Z_1,Z_2)^\top$ be a random vector with margins $F_1,F_2$ and joint cdf $F_{12}$ such that 
\begin{equation}
    \label{eq-copula}
    F_{12}(z_1,z_2) = C_{\bm Z}\{F_1(z_1),F_2(z_2)\},
\end{equation}
where $C_{\bm Z}$ is the {\it copula} function linking $Z_1$ and $Z_2$. A copula is simply a multivariate cdf with uniform $U(0,1)$ marginal cdfs, and \cite{Sklar1959} showed that the copula $C_{\bm Z}$ 
in (\ref{eq-copula}) is unique if the margins $F_1, F_2$ are continuous, and can be calculated as
\begin{equation*}
    C_{\bm Z}(u_1,u_2) = F_{12}\{F_1^{-1}(u_1),F_2^{-1}(u_2)\},\qquad 0 < u_1, u_2 < 1.
\end{equation*}

In this section, we shall study the tail properties of the copula $C_{\bm Z}$. In particular, using similar notation as in Section~\ref{intro}, but dropping the dependence on spatial lag $\hh$ for simplicity, we show that
$$
\lambda_U = \lim_{u \downarrow 0} \frac{-1+2u+C_{\bm Z}(1-u,1-u)}{u} > 0,
$$
i.e., the pair $(Z_1, Z_2)^{\top}$ is tail dependent in its upper tail, provided that the distance $\|\hh\|$ is sufficiently small. Moreover, we shall show in Section~\ref{subsec1.2} that $C_{12}$ has intermediate lower tail dependence, i.e.,
$$
C_{\bm Z}(u,u) \sim \ell(u)u^{\kappa_L}, \qquad u \downarrow 0,
$$
where  $\kappa_L \in (1, 2)$ is the lower tail order and $\ell(u)$ is a slowly varying function. In particular, this implies that $(Z_1, Z_2)^{\top}$ is tail-independent in its lower tail, i.e.,
$$
\lambda_L = \lim_{u \downarrow 0} \frac{C_{\bm Z}(u,u)}{u} = 0.
$$

Let $C^n_{\bm Z}$ be the copula of the vector of componentwise maxima from i.i.d.  copies $\bm Z_i=(Z_{i1},Z_{i2})^\top$ of $\bm Z$, $i=1,\ldots,n$, i.e., $\bm M_n=(M_{n1},M_{n2})^\top$ with $M_{nj}=\max(Z_{1j}, \ldots, Z_{nj})$, $j=1,2$. Extreme-value copulas, denoted $C_{\rm EV}$, describe the class of dependence structures that arise as limits of $\bm M_n$ (when properly renormalized), i.e.,
\begin{equation}\label{eq:EVC}
C_{\rm EV}(u_1,u_2)=\lim_{n\to\infty} C^n_{\bm Z}(u_1^{1/n},u_2^{1/n}),\qquad (u_1,u_2)^\top\in[0,1]^2.
\end{equation}
It can be shown that extreme-value copulas are such that for any $k=1,2,\ldots,$ one has $C_{\rm EV}(u_1,u_2)=C_{\rm EV}^k(u_1^{1/k},u_2^{1/k})$, $(u_1,u_2)^\top\in[0,1]^2$, and they can be characterized as
\begin{equation}\label{eq:stabletail}
C_{\rm EV}(u_1,u_2)=\exp\{-\ell_{\bm Z}(-\log u_1,-\log u_2)\},\qquad (u_1,u_2)^\top\in[0,1]^2,
\end{equation}
where $\ell_{\bm Z}$ is called the \emph{stable (upper) tail-dependence function} and completely determines the limiting extremal dependence structure of $\bm Z$ in the upper tail. From \eqref{eq:EVC} and \eqref{eq:stabletail}, the stable tail-dependence function can be expressed as the limit $\ell_{\bm Z}(w_1,w_2)=\lim_{n\to\infty}n\{1-C_{\bm Z}(1-w_1/n,1-w_2/n)\}$, and
the next proposition gives the stable tail dependence function of the process \eqref{eq2}.
\begin{prop} \label{prop2} \rm Assume that $Z(\ss)$ is defined as in \eqref{eq2} such that $\sup_{\ss^* \in \AA(\ss)} W(\d\ss^*)$ has the unit Fr\'echet distribution if $|\AA(\ss)| = 1$.  
The stable tail-dependence function of the random vector $\bm Z = (Z_{1}, Z_{2})^{\top} = (Z(\ss_1), Z(\ss_2))^{\top}$ is 
\begin{multline}
\ell_{\bm Z}(w_1, w_2) = w_1 \frac{\int_{(0,\infty)^3} x_1h_1(\xx)g(\xx) \d\xx}{\int_{(0,\infty)^3}(x_1+x_{12})h_1(\xx)g(\xx)\d\xx} +  w_2 \frac{\int_{(0,\infty)^3} x_2h_2(\xx)g(\xx) \d\xx}{\int_{(0,\infty)^3}(x_2+x_{12})h_2(\xx)g(\xx)\d\xx}\\
+ \int_{(0,\infty)^3}x_{12}\max \left\{\frac{w_1h_1(\xx) g(\xx)}{\int_{(0,\infty)^3}(x_1+x_{12})h_1(\xx)g(\xx)\d\xx}, \frac{w_2h_2(\xx)g(\xx)}{\int_{(0,\infty)^3}(x_2+x_{12})h_2(\xx)g(\xx)\d\xx}\right\} \d\xx\,, \quad w_1, w_2 > 0.\label{eq-ell1}
\end{multline}
\end{prop}

\noindent
\textbf{Proof:} By definition of the stable tail-dependence function, we need to compute the limit $\ell_{\bm Z}(w_1,w_2)=\lim_{n\to\infty} n\{1-p_n(w_1,w_2)\}$, where  $p_n(w_1, w_2) = F_{12}(F_1^{-1}(1-w_1/n), F_2^{-1}(1-w_2/n))$ as $n \to \infty$. From (\ref{eq-bivcdf}), the marginal cdf of $Z_i$ is
\begin{eqnarray*}
	F_i(z) &=&  \int_{(0,\infty)^3} \exp{\left\{-\frac{(x_i+x_{12})h_i(\xx)}{z}\right\}} g(\xx) \d\xx\\
	&=& 1 - z^{-1} \int_{(0,\infty)^3} (x_i+x_{12})h_i(\xx)  g(\xx) \d\xx + o(z^{-1}), \quad z \to \infty,
\end{eqnarray*}
and therefore $$F_i(z_i^*) = 1-\frac{w_i}{n} + o(n^{-1}), \quad \text{with } z_i^* = \frac{n}{w_i} \int_{(0,\infty)^3} (x_i+x_{12})h_i(\xx)  g(\xx) \d\xx.$$

This implies using (\ref{eq-bivcdf}), that
$$
p_n(w_1, w_2) = F_{12}(z_1^*, z_2^*) + o(n^{-1}) = 1-\frac{1}{n}\int_{(0,\infty)^3}P(w_1,w_2;\xx)g(\xx)\d\xx + o(n^{-1}),$$ 
where we use continuity of $F_{12}(z_1,z_2)$ to get the first equality and 
$$P(w_1,w_2;\xx)= w_1h_1^*(\xx)+w_2h_2^*(\xx) + \max\left\{w_1 \tilde h_1(\xx), w_2 \tilde h_2(\xx)\right\} $$
with
$$
h_i^*(\xx)=\frac{x_ih_i(\xx)}{\int_{(0,\infty)^3}(x_i+x_{12})h_i(\xx)g(\xx)\d\xx}\,, \quad \tilde h_i(\xx)=\frac{x_{12}h_i(\xx)}{\int_{(0,\infty)^3}(x_i+x_{12})h_i(\xx)g(\xx)\d\xx}\,. \\
$$
Therefore, we get that $\ell_{\bm Z}(w_1,w_2) = \lim_{n \to \infty}n\{1-p_n(w_1,w_2)\} =  \int_{(0,\infty)^3}P(w_1,w_2;\xx)g(\xx)\d\xx$. \hfill $\Box$

\begin{corol} \label{corol1} \rm Under the assumptions of proposition \ref{prop2}, $C_{\bm Z}$ has upper tail dependence with
$$
\lambda_U = 2 - \ell_{\bm Z}(1,1) = \int_{(0,\infty)^3}x_{12}\min \left\{\frac{h_1(\xx)g(\xx)}{\int_{(0,\infty)^3}(x_1+x_{12})h_1(\xx)g(\xx)\d\xx}, \frac{h_2(\xx)g(\xx)}{\int_{(0,\infty)^3}(x_2+x_{12})h_2(\xx)g(\xx)\d\xx}\right\} \d\xx.
$$
It implies that $\lambda_U = 0$ if and only if $g(\xx) = 0$ for $x_{12} > 0$, except for a set of measure zero. It follows that $\lambda_U = 0$ if $\Pr(\AA_{12}) = 0$, that is, the random sets $\AA(\ss_1)$ and $\AA(\ss_2)$ do not overlap with probability one. 
\end{corol}

\subsection{Simplified special case}
\label{subsec1.2}

We now consider a special case of the process \eqref{eq2} which makes inference simpler and still allows for high flexibility when modeling stationary and isotropic data with rough spatial fields. For simplicity, we here assume $q=2$.  We consider random sets that are disks with a random radius whose dependence structure is driven by a spatially-correlated Gaussian copula. Specifically, we make the following assumptions. Let $h(z) = 1/z$ and $\AA(\ss) = \{\ss^*: ||\ss-\ss^*|| < R(\ss)\}$, where $R(\ss)$ is a trans-Gaussian spatial process with uniform $U(r_L, r_U)$ marginals and some isotropic correlation function $\rho_R(\cdot)$. The process \eqref{eq2} can thus now be written as:
\begin{equation}
\label{eq2s}
Z(\ss) = \frac{1}{\pi R(\ss)^2}\sup_{\ss^*:  ||\ss-\ss^*|| < R(\ss)} W(\d\ss^*), \quad \ss \in \mathbb{R}^2.
\end{equation}

Note that $Z(\ss_{1})$ and $Z(\ss_{2})$ are independent if $h = ||\ss_{1} - \ss_{2}|| > r_U$ since $\AA(\ss_{1}) \cap \AA(\ss_{2}) = \emptyset$ in this case. The parameter $r_U$ therefore controls the dependence range for the process $Z(\ss)$. On the other hand, the parameter $r_L$ has an effect on the smoothness of the field, as we shall see.

Note that while we here assume that $R(\boldsymbol{s})$ is an isotropic process, all the results presented below can be easily extended to the general case of a nonstationary process $R(\ss)$. Let $g_R(r_{1}, r_{2}; \delta)$ be the continuous joint density of $(R(\ss_{1}), R(\ss_{2}))^{\top}$. For $i=1,2$, let $x_i + x_{12} = \pi r_i^2$ and $\delta_i = x_{12}/(\pi r_i^2)$, so that $x_i = \pi r_i^2 (1-\delta_i)$ and $x_{12} = \pi r_i^2\delta_i$, we can then rewrite (\ref{eq-ell1}) as
\begin{equation*}
\ell(w_1, w_2) = w_1 + w_2 -
\int_{(0,\infty)^2} \min\left\{w_1\delta_1(r_1,r_2;h), w_2\delta_2(r_1,r_2;h)\right\} g_R(r_1,r_2;h)\d r_1 \d r_2,
\end{equation*}
where $\delta_i \equiv \delta_i(r_1,r_2;h) = A_{12} h(A_i + A_{12}) \equiv A_{12}(r_1, r_2;h)/(\pi r_i^2)$, $i=1,2$, with
\begin{eqnarray*}
A_{12}(r_1,r_2;h) =& |\AA_{12}|, \quad  &\AA_{12} = \{\ss^* \in \mathbb{R}^2: ||\ss_1 - \ss^*|| < r_1, ||\ss_2 - \ss^*|| < r_2\},\\
A_{i}(r_1,r_2;h) =& |\AA_{i}| = \pi r_i^2, \quad  &\AA_{i}\ \, = \{\ss^* \in \mathbb{R}^2: ||\ss_i - \ss^*|| < r_i\}.
\end{eqnarray*}

It can be shown that
$$A_{12}(r_1,r_2;h) = r_1^2\{\phi_1 - 0.5\sin(2\phi_1)\} + r_2^2\{\phi_1 - 0.5\sin(2\phi_2)\},$$ $$\phi_1 = \arccos\left(\frac{h^2 + r_1^2 - r_2^2}{2h r_1}\right), \quad \phi_2 = \arccos\left(\frac{h^2 + r_2^2 - r_1^2}{2h r_2}\right)\,,$$ if $|r_1 - r_2| < h$, and $A_{12}(r_1, r_2;h) = \min\{\pi r_1^2, \pi r_2^2\}$ if $|r_1 - r_2| \geq h$.

From (\ref{eq2s}), the process $Z(\ss)$ has standard Fr\'echet marginals with $F_i(z) = \Pr\{Z(\ss) \leq z\} = \exp(-1/z)$, $z > 0$. Since $x_i = \pi r_i^2\left\{1-\delta_i(r_1,r_2;h)\right\}$ and $x_{12} = \pi r_i^2\delta_i(r_1,r_2;h)$, we find from (\ref{eq-bivcdf}) that the copula of $\ZZ = (Z_1, Z_2)^{\top}$ linking $Z_1=Z(\ss_{1})$ and $Z_2=Z(\ss_{2})$ is 
\begin{equation}
\label{eq-copcdf}
C_{\ZZ}(u_1, u_2; h) = \int_{(0,\infty)^2} u_1^{1-\delta_1(r_1,r_2;h)}u_2^{1-\delta_2(r_1,r_2;h)}\min\{u_1^{\delta_1(r_1,r_2;h)}, u_2^{\delta_2(r_1,r_2;h)}\}g_R(r_1,r_2;h)\d r_1\d r_2.
\end{equation}
The resulting copula is therefore a mixture of Marshall--Olkin copulas \citep{Marshall.Olkin1967}, mixed over its parameters $\delta_1$ and $\delta_2$, and the resulting process $Z(\ss)$ is a max-mixture of indicator kernel spatial processes. From (\ref{eq-copcdf}), by exploiting the properties of Marshall--Olkin copulas, we easily find that:
\begin{equation}
\label{eq-lam-srho}
\begin{split}
	\lambda_U(h) &= \int_{(0,\infty)^2}\min\{\delta_1(r_1,r_2;h), \delta_2(r_1,r_2;h)\} g_R(r_1,r_2;h)\d r_1\d r_2,\\
	S_{\rho}(h) &= \int_{(0,\infty)^2} 3\{2/\delta_1(r_1,r_2;h) + 2/\delta_2(r_1,r_2;h) - 1\}^{-1} g_R(r_1,r_2;h)\d r_1\d r_2,
\end{split}
\end{equation}
where $S_{\rho}(h)$ denotes the Spearman's correlation coefficient of the copula $C_{\bm Z}$, which is a measure of dependence in the bulk of the distribution. Note that $C_{\bm Z}$ has a continuous density $c_{\bm Z}(u_1,u_2)$ unless $r_L = r_U = r > 0$ (when $C_{\bm Z}$ is the Marshall--Olkin copula with parameters $\delta_1(r,r;h)$ and $\delta_2(r,r;h)$).
 
Further note that 
$$
C_{\bm Z}(u,u;h) = \int_{(0,\infty)^2} u^{2-\min\{\delta_1(r_1,r_2;h), \delta_2(r_1,r_2;h)\}} g_R(r_1,r_2;h)\d r_1\d r_2.
$$
Since $\sup_{r_L < r_1,r_2 < r_U} \min\{\delta_1(r_1,r_2;h), \delta_2(r_1,r_2;h)\} = \delta_1(r_U,r_U;h)$ and $g_R(r_1,r_2;h)$ is a continuous function, this implies that 
$C_{\bm Z}(u,u;h) \sim K_Lu^{2-\delta_1(r_U,r_U;h)}$ as $u \to 0$, where $K_L$ is a positive constant. Therefore, in the lower tail, the copula $C_{\bm Z}$ has intermediate tail-dependence when $h > 0$, and the lower tail order is given by $\kappa_L = 2 -\delta_1(r_U,r_U;h)$. This means that the proposed model is (locally) tail-dependent in the upper tail with positive $\lambda_U(h)$ and tail-independent in the lower tail with $\lambda_L(h)=0$, yet with some flexibility in capturing the strength of lower tail dependence.
 
Now, let us investigate the local behavior of the proposed process, for small distances $||\ss_1 - \ss_2||$. 
\begin{prop}
    \label{prop1}
    Assume that $\Pr\{|R(\ss_1) - R(\ss_2)| > h\} > K_0 > 0$ as $h = ||\ss_1 - \ss_2|| \to 0$, then $\lambda_U(h) \leq 1 - K_0 h (2r_L+h)/r_U^2$.
\end{prop}

\noindent
\textbf{Proof:}
We find that
\begin{eqnarray*}
\lambda_U(h)  &\leq& \int_{|r_1-r_2| < h} g_R(r_1,r_2;h) \d r_1 \d r_2\\
&& +  \int_{r_2 > r_1 + h} \frac{r_1^2}{r_2^2}\,g_R(r_1,r_2;h) \d r_1 \d r_2 + \int_{r_1 > r_2 + h} \frac{r_2^2}{r_1^2}\,g_R(r_1,r_2;h) \d r_1 \d r_2\\
&& =  1 -\int_{r_2 > r_1 + h}  \left(1-\frac{r_1^2}{r_2^2}\right)\,g_R(r_1,r_2;h) \d r_1 \d r_2 - \int_{r_1 > r_2 + h}  \left(1-\frac{r_2^2}{r_1^2}\right)\,g_R(r_1,r_2;h)\d r_1 \d r_2.
\end{eqnarray*}
Since $$1-\frac{r_1^2}{r_2^2} = \frac{(r_2 - r_1)(r_1 + r_2)}{r_2^2} \geq \frac{h (2r_L + h)}{r_U^2} \qquad \text{if } r_2 > r_1 + h,$$
we find that $$\lambda_U(h) \leq 1 - \frac{h (2r_L + h)}{r_U^2} \int_{|r_1 - r_2| > h}g_R(r_1,r_2)\d r_1 \d r_2 \leq 1 - K_0\frac{h (2r_L + h)}{r_U^2},$$
which concludes the proof. \hfill $\Box$

A similar result holds for $S_{\rho}(h)$. This implies that the process $Z(\ss)$ can therefore be used to model data with rough realizations at extreme levels where $1-\lambda_U(h)=O(h^\alpha)$ with $\alpha\leq 1$, and similarly for $S_\rho(h)$. 

\begin{figure}[t!]
	\begin{center}
		\includegraphics[width=0.48\linewidth]{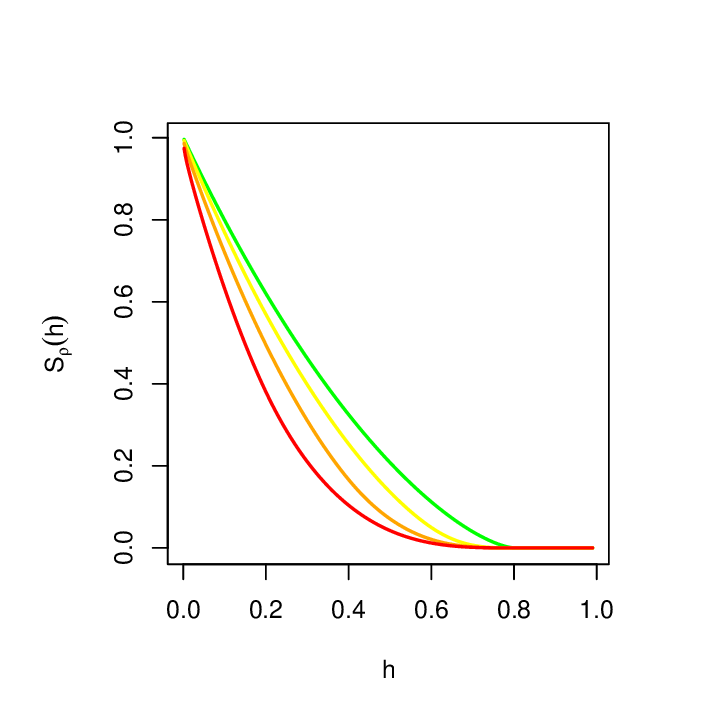}\hspace{-10mm}
		\includegraphics[width=0.48\linewidth]{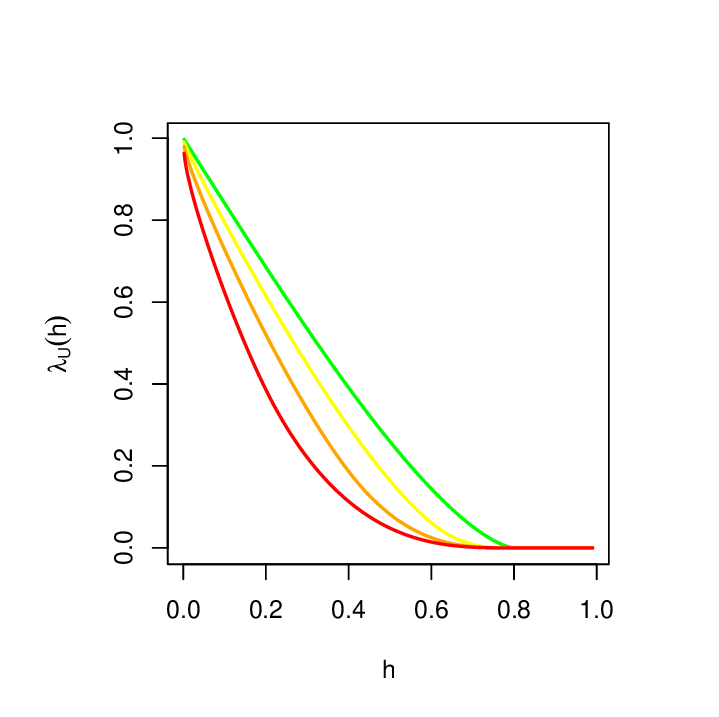}
		\caption{{\footnotesize {Spearman's $\rho$, $S_{\rho}(h)$ (left) and tail dependence coefficient, $\lambda_U(h)$ (right) as in (\ref{eq-lam-srho}) calculated for $r_U = 0.4$ and $r_L = 0.1, 0.2, 0.3, 0.4$ (red, orange, yellow, green lines, respectively). The density $g_R$ is given by the Gaussian copula with $U(r_L, r_U)$ marginals and the correlation $\rho_R(h) = \exp(-h)$. }}}
		\label{fig-srho-lam}
	\end{center}
\end{figure}

\begin{figure}[t!]
	\begin{center}
		\includegraphics[width=2.2in]{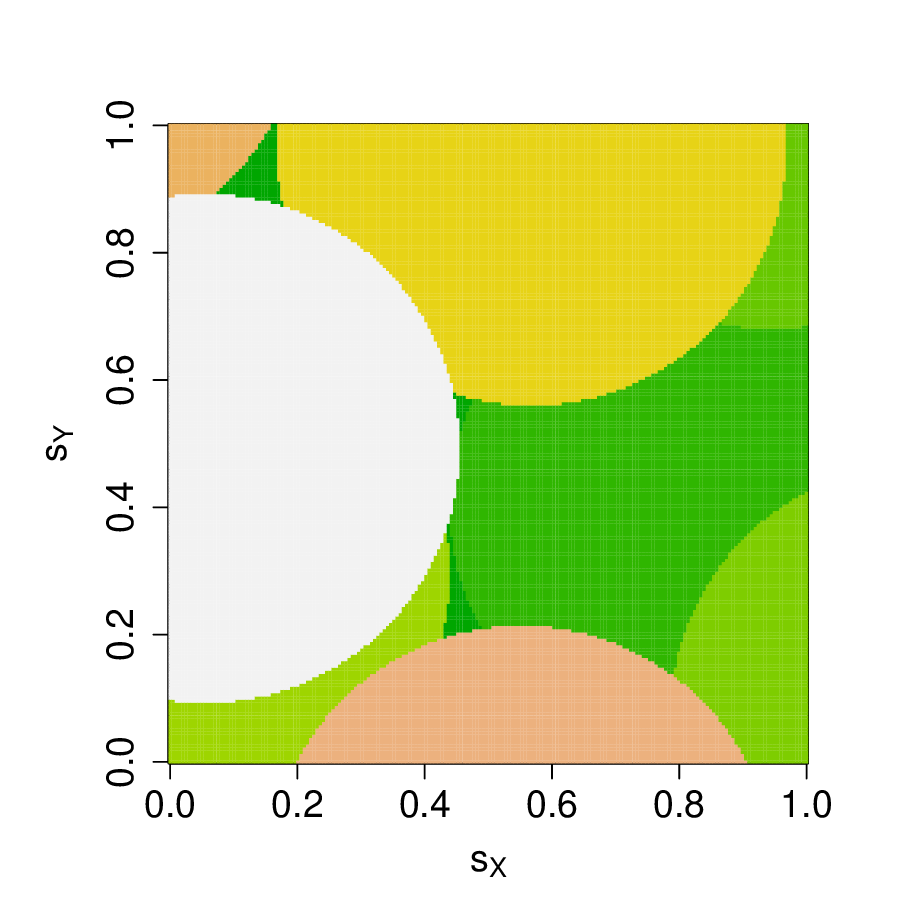}
		\hspace{-0.5cm}
		\includegraphics[width=2.2in]{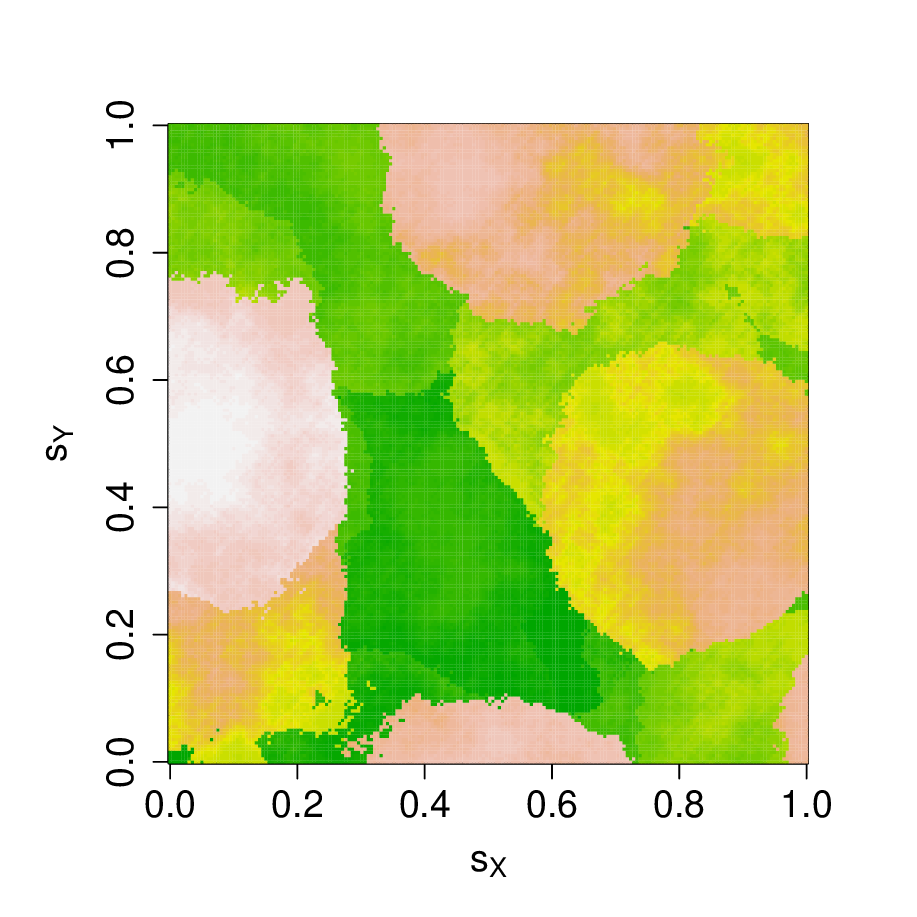}
		\hspace{-0.5cm}
		\includegraphics[width=2.2in]{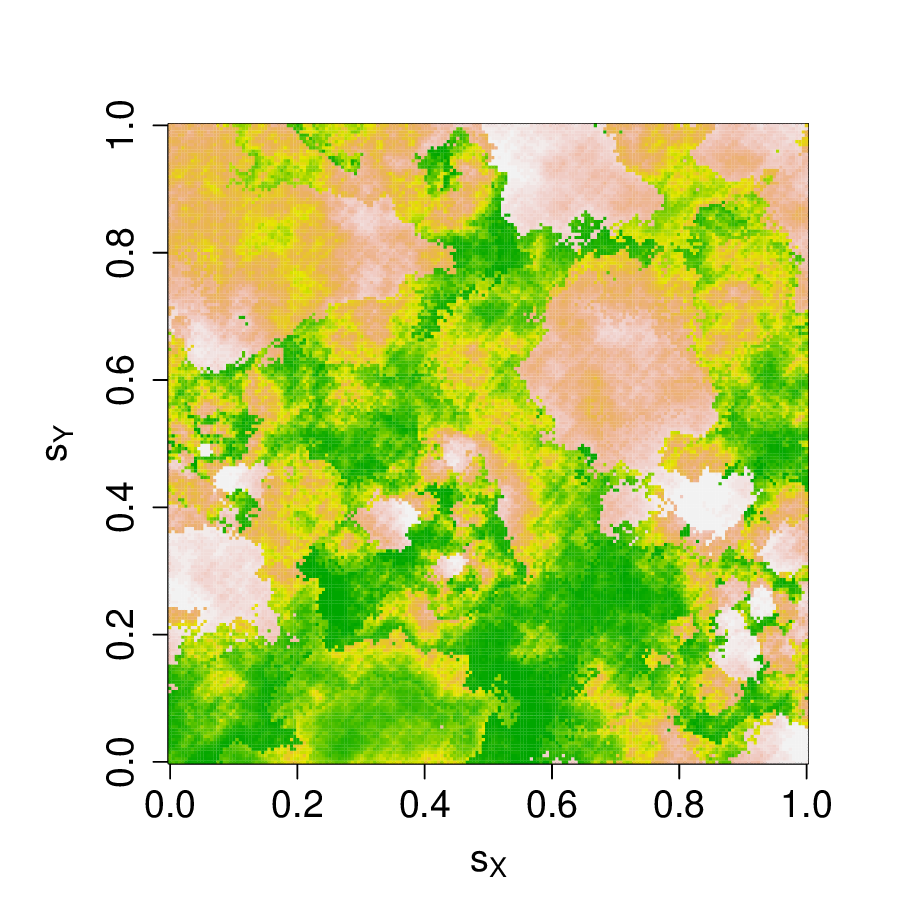}
		\caption{{\footnotesize Realizations of the process $Z(\ss)$ as defined in (\ref{eq2s}) with $U(0,1)$ marginals. We assume $R(\ss)$ is a Gaussian process with $U(r_L,r_U)$ marginals, exponential covariance function $\rho_R(h) = \exp(-h)$ and $r_U = 0.4$, and $r_L = 0.4$ (left), $r_L = 0.2$ (middle) and $r_L = 0.0$ (right).}}
		\label{Fig1}
	\end{center}
\end{figure}

Since $A_{12}(r_1,r_2;h) = 0$ if $r_1 + r_2 \leq h$, we have that $Z(\ss_1)$ and $Z(\ss_2)$ are independent if and only if $r_1 + r_2 \leq r_U$. Hence, while the parameter $r_U$ controls the range of dependence for the process $Z(\ss)$, $r_L$ controls its smoothness behavior. Figure~\ref{fig-srho-lam} shows $S_{\rho}(\delta)$ and $\lambda_U(\delta)$ given in (\ref{eq-lam-srho}) computed for fixed $r_U=0.4$ and different values of $r_L$, assuming that $R(\ss)$ is a trans-Gaussian process with $U(r_L, r_U)$ marginals and exponential correlation function $\rho_R(h) = \exp(-h)$. We can see that smaller values of $r_L$ corresponds to a faster rate of decay of $S_{\rho}(h)$ and $\lambda_U(h)$, which results in a process with rougher realizations. Figure~\ref{Fig1} shows simulations of this process for different parameter values, indeed with rougher realizations for smaller values of $r_L$. For the remainder of the paper, we consider the simplified definition of the process $Z(\ss)$ as defined in (\ref{eq2s}).

\subsection{Model extension}
\label{subsec1.3}

The proposed model (\ref{eq2s}) allows for high flexibility in the joint tail; however, it lacks flexibility in the bulk of the distribution. Figure~\ref{fig-srho-lam} indicates that the Speraman's rho coefficient, $S_{\rho}(h)$, and the upper tail dependence coefficient, $\lambda_U(h)$, indeed follow very similar patterns. In particular, $\lambda_U(h) = 0$ if and only if the two respective realizations of the process $Z(\ss)$ in \eqref{eq2} are exactly independent. However, strong overall dependence and asymptotic tail-independence can be observed in many applications, and in this section we extend the model \eqref{eq2} to construct a new process that allows for this type of dependence structure.  

We consider an extension that allows for upper tail-dependence at smaller distances and tail-independence with strong bulk dependence at larger distances.

Let $Z(\ss)$ be the process as defined in Section~\ref{subsec1.2}, and $Y(\ss)$ be a spatial process with marginal cdf $F_Y$. 
We define the max-mixture process 
\begin{equation}
\label{eq-convproc}
\tilde Z(\ss) = \max\{q Z(\ss), (1-q)Y(\ss)\}\,, \quad 0 < q \leq 1.
\end{equation}
This construction is similar to the hybrid spatial dependence model introduced by \cite{Wadsworth.Tawn:2012b} but here with different marginal distributions for $Z(\ss)$ and $Y(\ss)$. The main goal is to find a simple process $Y(\ss)$ that does not affect the tail dependence of the original process $Z(\ss)$, but that allows for greater flexibility at subasymptotic levels. Here, we  assume that the marginal distribution satisfies $\bar F_Y(z) \sim z^{-\beta}$, $\beta > 0$ as $z \to \infty$.

By construction, the marginal survival function of process $\tilde Z(\ss)$ in this case is
$$
\bar{\tilde F}(z) = \Pr\{\tilde Z(\ss) > z\} = 1-e^{-q/z}\cdot\left\{1-\bar F_Y\left(\frac{z}{1-q}\right)\right\} = \frac{q}{z} +  \frac{(1-q)^{\beta}}{z^{\beta}} + o(z^{-\beta}) + O(z^{-2}), \quad z \to \infty. 
$$
Let $\tilde C$, $C_{\bm Z}$, and $C_{\bm Y}$ be the copula linking $(\tilde Z(\ss_1), \tilde Z(\ss_2))^{\top}$, $(Z(\ss_1), Z(\ss_2))^{\top}$, and $(Y(\ss_1), Y(\ss_2))^{\top}$, respectively, with $C_{\bm Z}$ defined in (\ref{eq-copcdf}). Let $\tilde\lambda_U$, $\lambda_U^{\bm Z}$, $\lambda_U^{\bm Y}$ be the upper tail-dependence coefficient of $\tilde C$, $C_{\bm Z}$, $C_{\bm Y}$, and let  $\tilde\kappa_U$, $\kappa_U^{\bm Y}$ ($\tilde\kappa_L$, $\kappa_L^{\bm Y}$) be the upper (lower) tail order of $\tilde C$, $C_{\bm Y}$, respectively, where we here drop the dependence on distance $h$ for simplicity. The following cases are possible with model~(\ref{eq-convproc}):
\begin{itemize}
    \item If both $C_{\bm Z}$ and $C_{\bm Y}$ have upper tail-dependence, then it follows that $$\Pr(\tilde Z(\ss_1) > z, \tilde Z(\ss_2) > z) \sim \lambda_U^{\bm Z}\frac{q}{z} + \lambda_U^{\bm Y}\frac{(1-q)^{\beta}}{z^{\beta}} + o(z^{-\beta}) + O(z^{-2}).$$ In particular, if $\beta < 1$, then $\tilde\lambda_U = \lambda_U^{\bm Y}$. If $\beta = 1$, then $\tilde\lambda_U = q\lambda_U^{\bm Z} + (1-q)\lambda_U^{\bm Y}$, and if $\beta > 1$, then $\tilde\lambda_U = \lambda_U^{\bm Z}$. This implies that if $\beta > 1$, the new process $\tilde Z(\ss)$ has the same tail-dependence structure as the max-convolution process $Z(\ss)$. 
    \item If $C_{\bm Z}$ has upper tail-dependence, and $C_{\bm Y}$ does not, then it follows that $$\Pr(\tilde Z(\ss_1) > z, \tilde Z(\ss_2) > z) \sim \lambda_U^{\bm Z}\frac{q}{z}  + \ell_{\bm Y}(z)\frac{1}{z^{\beta \kappa_U^{\bm Y}}} +o(z^{-\beta\kappa_U^{\bm Y}}) + O(z^{-2}),$$ where $\ell_{\bm Y}(z)$ is a slowly varying function. In this case, if $\beta < 1$, then $\tilde\lambda_U = 0$ and $\tilde\kappa_U = \min(1/\beta, \kappa_U^{\bm Y})$. If $\beta = 1$, then $\tilde\lambda_U = q\lambda_U^{\bm Z}$, and if $\beta > 1$, then $\tilde\lambda_U = \lambda_U^{\bm Z}$, which means that mixing the original process $Z(\ss)$ with a tail independent process $Y(\ss)$ does not affect the tail dependence if the marginals of the latter process have lighter tails. 
    \item If $C_{\bm Z}$ does not have tail dependence (and so $C_{\bm Z}$ is the independence copula), and $C_{\bm Y}$ does, then it follows that $$\Pr(\tilde Z(\ss_1) > z, \tilde Z(\ss_2) > z) \sim \lambda_U^{\bm Y}\frac{(1-q)^{\beta}}{z^{\beta}} + o(z^{-\beta}) + O(z^{-2}).$$ If $\beta < 1$, then $\tilde\lambda_U = \lambda_U^{\bm Y}$. On the other hand, if $\beta=1$, then $\tilde\lambda_U = (1-q)\lambda_U^{\bm Y}$, and if $\beta > 1$, then $\tilde\lambda_U = 0$ and $\tilde\kappa_U = \min(\beta,2)$. In the last case, the tail index $\beta$ controls the strength of dependence of the process $\tilde Z(\ss)$ at subasymptotic levels. We have that $\tilde \kappa_U \leq 2$, so that the process cannot capture negative association in the upper tail.
    \item If both $C_{\bm Z}$ and $C_{\bm Y}$ do not have upper tail dependence, then it follows that $$\Pr(\tilde Z(\ss_1) > z, \tilde Z(\ss_2) > z) \sim  \ell_{\bm Y}z^{-\beta\kappa_U^{\bm Y}} + o(z^{-\beta\kappa_U^{\bm Y}}) + O(z^{-2}),$$ where $\ell_{\bm Y}(z)$ is a slowly varying function. Here, $\tilde\lambda_U = 0$, $\tilde\kappa_U = \min(\kappa_U^{\bm Y},2/\beta)$ if $\beta \leq 1$, and $\tilde\kappa_U = \min(\beta\kappa_U^{\bm Y},2)$ if $\beta > 1$.
\end{itemize}

The above results imply that mixing $Z(\ss)$ with a process $Y(\ss)$ with lighter tails ($\beta > 1$) does not affect the upper tail-dependence properties of the new process $\tilde Z(\ss)$. At the same time, if $Y(\ss)$ does not have tail-dependence (e.g., it is a marginally transformed Gaussian process), the strength of dependence as measured by $\tilde\kappa_U = \min(\beta\kappa_U^{\bm Y}, 2)$ for $\beta > 1$ can be quite weak. In particular, the proposed process $\tilde Z(\ss)$ has upper tail-dependence at small distances, and intermediate dependence or tail quadrant independence with $1 < \tilde\kappa_U \leq 2$ at large distances if $Y(\ss)$ is based on a Gaussian process. Thus, this process can effectively control the strength of dependence in the bulk of the joint distribution even at large distances.

Note that the above results with $\beta > 1$ can also be extended to processes $Y(\ss)$ with marginals with lighter tails, such as the standard normal marginals which can be considered as a limiting case with $\beta \to \infty$. 

\begin{figure}[t!]
	\begin{center}
		\includegraphics[width=0.5\linewidth]{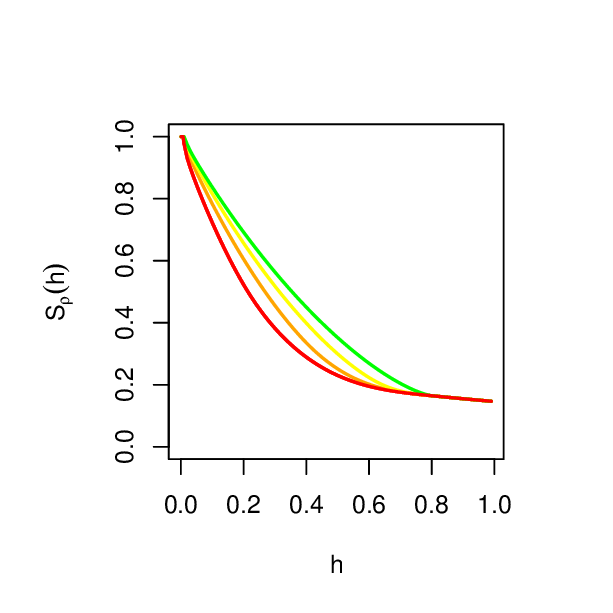}\hspace{-10mm}
		\includegraphics[width=0.5\linewidth]{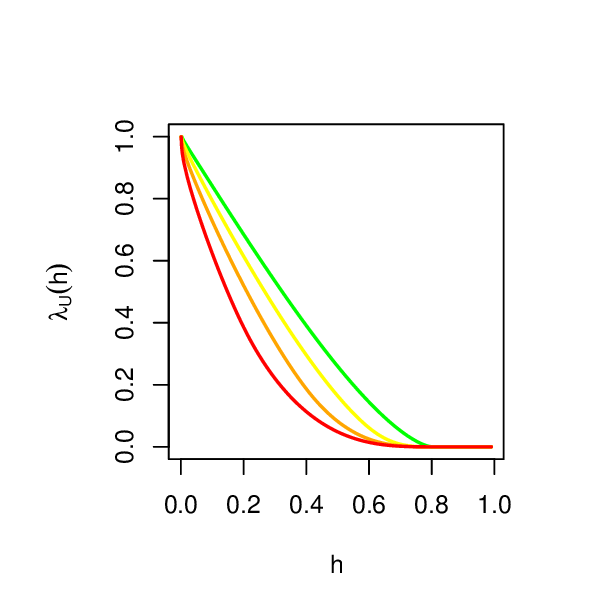}
		\caption{{\footnotesize {Spearman's $\rho$, $S_{\rho}(h)$ (left) and tail dependence coefficient, $\lambda_U(h)$ (right) calculated for the process $\tilde Z(\ss)$ defined in (\ref{eq-convproc}) with $q=0.2$, $r_U = 0.4$ and $r_L = 0.1, 0.2, 0.3, 0.4$ (red, orange, yellow, green lines, respectively). We assume $\rho_R(h) = \exp(-h)$, and $Y(\ss)$ is a Gaussian process with $N(0,1)$ marginals that has the exponential covariance function  $\rho_Y(h) = \exp(-h/2)$. }}}
		\label{fig2-srho-lam}
	\end{center}
\end{figure}

Figure~\ref{fig2-srho-lam} shows Spearman's $\rho$, $S_{\rho}(h)$, and the upper tail dependence coefficient, $\lambda_U(h)$, for the process $\tilde Z(\ss)$ defined in (\ref{eq-convproc}) with $q = 0.2$ computed for different values of $r_L$, assuming $R(\ss)$ is a trans-Gaussian process with $U(r_L, r_U)$ marginals and exponential covariance function $\rho_R(h) = \exp(-h)$, and $Y(\ss)$ is a standard Gaussian process with exponential covariance function $\rho_Y(h) = \exp(-h/2)$. We can see that $\lambda_U(h)$ is indeed the same as before (recall Figure~\ref{fig-srho-lam}), while the range of overall dependence is now controlled by the process $Y(\ss)$, and for the selected parameters, Spearman's $\rho$ is larger compared to Figure~\ref{fig-srho-lam}.

\begin{figure}[t!]
	\begin{center}
		\includegraphics[width=2.2in]{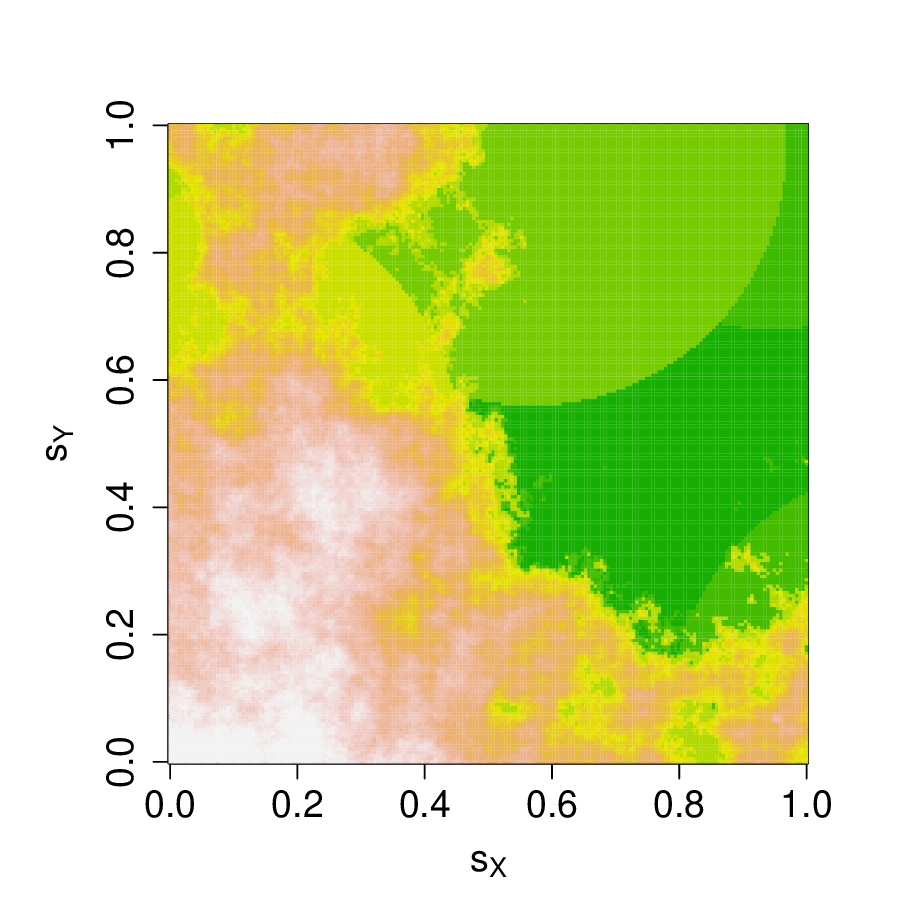}
		\hspace{-0.5cm}
		\includegraphics[width=2.2in]{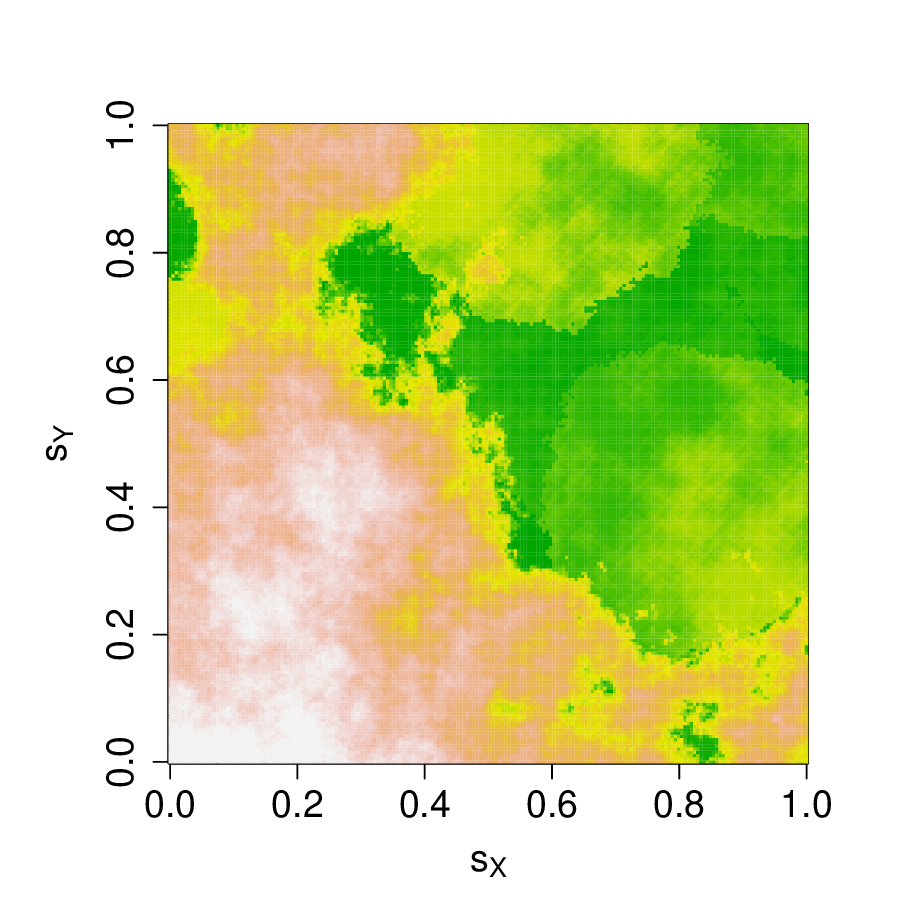}
		\hspace{-0.5cm}
		\includegraphics[width=2.2in]{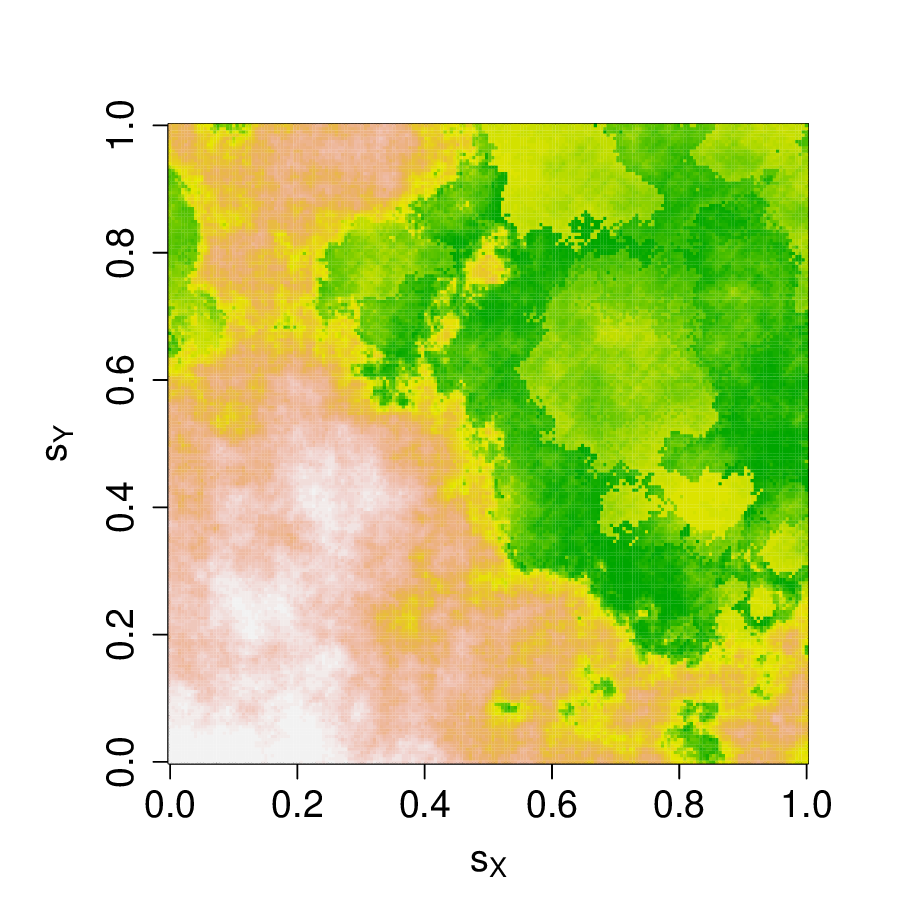}\\
		\includegraphics[width=2.2in]{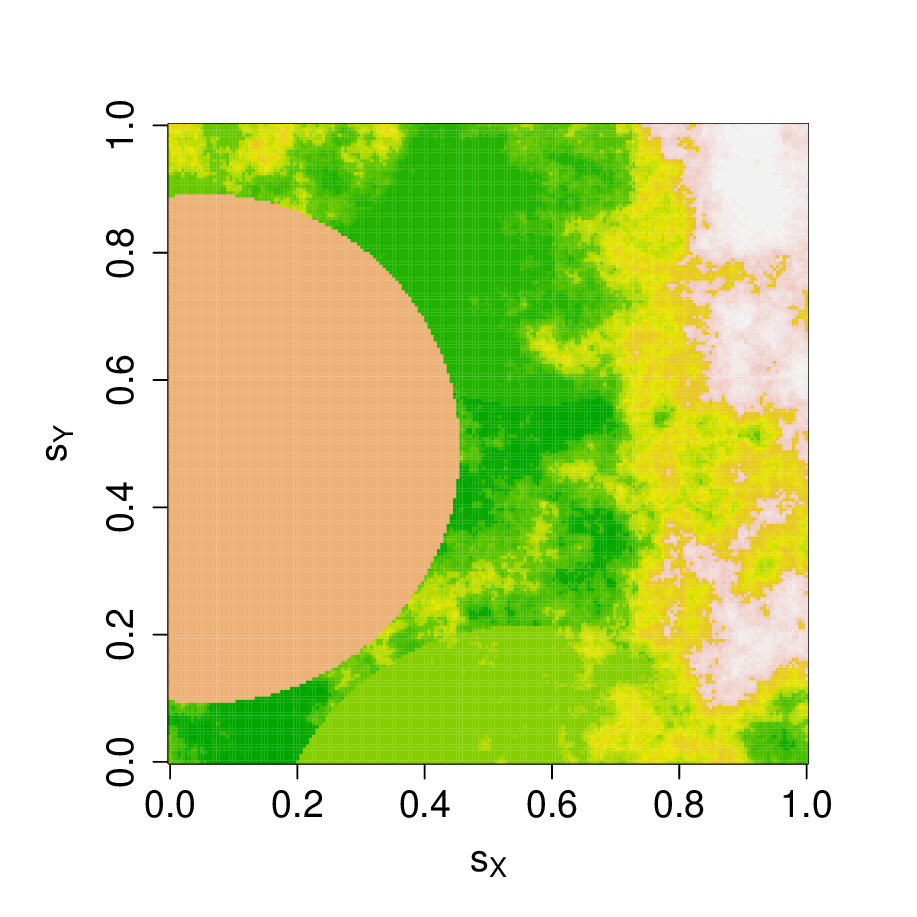}
		\hspace{-0.5cm}
		\includegraphics[width=2.2in]{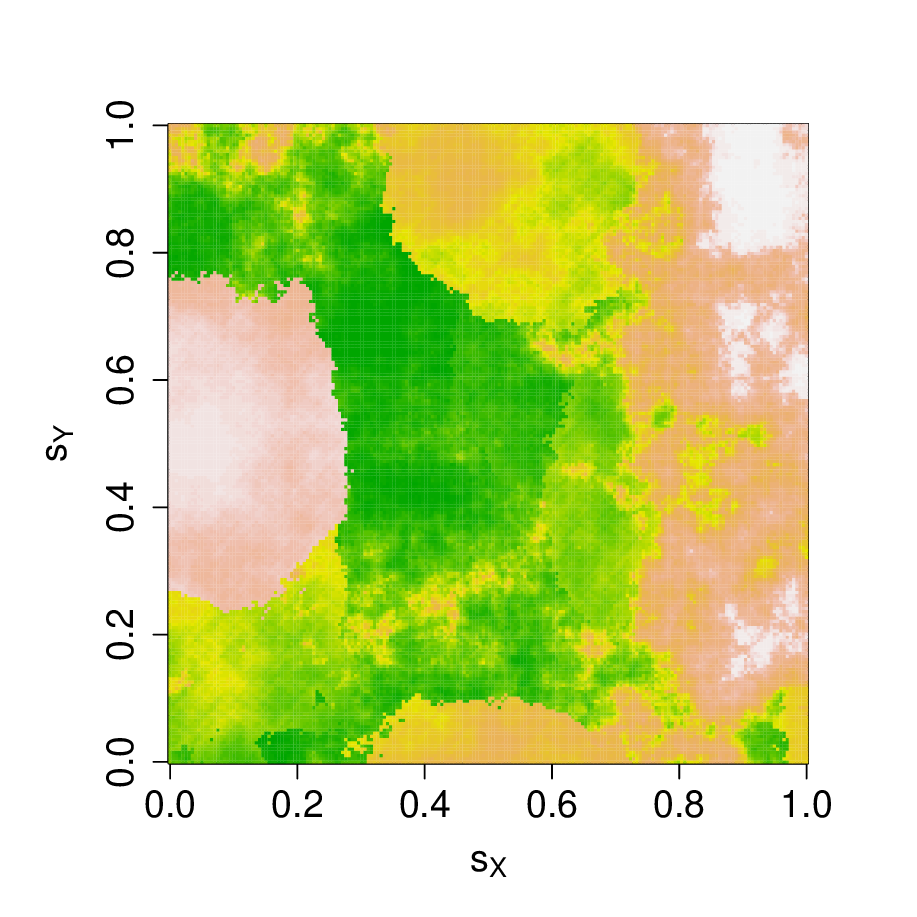}
		\hspace{-0.5cm}
		\includegraphics[width=2.2in]{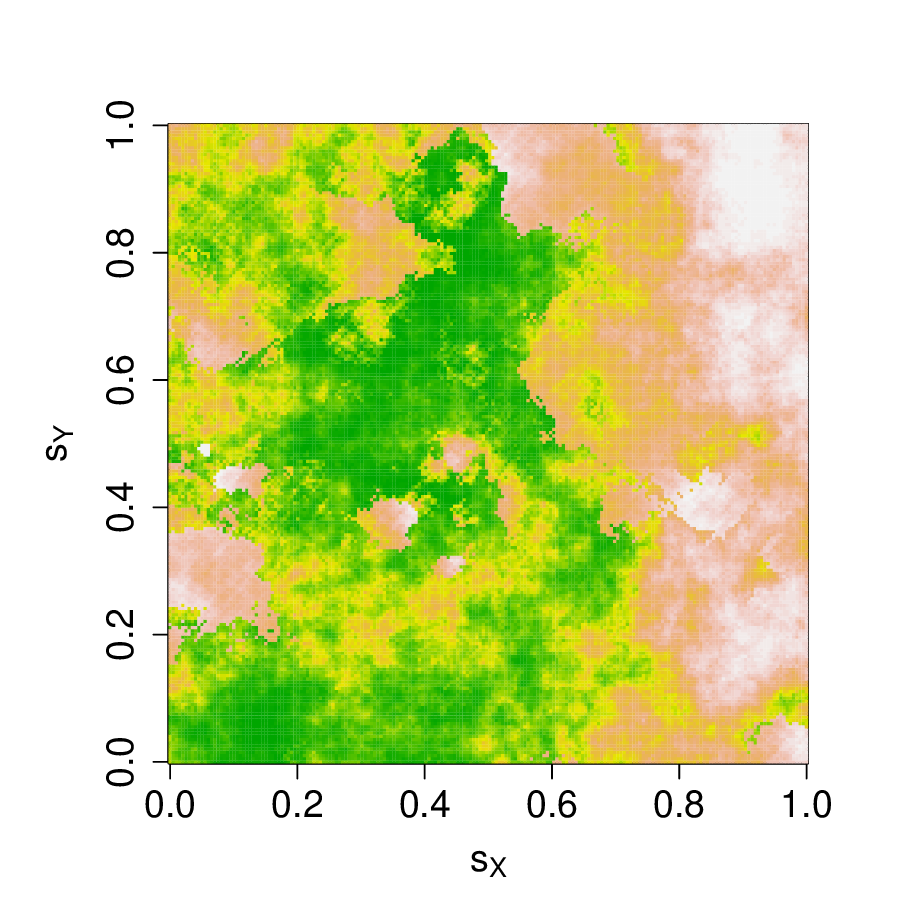}
		\caption{{\footnotesize Realizations of the max-mixture process $\tilde Z(\ss)$ as defined in (\ref{eq-convproc}) with $q=2$, $U(0,1)$ marginals, exponential covariance function $\rho_R(h) = \exp(-h)$ and $r_U = 0.4$, and $r_L = 0.4$ (left), $r_L = 0.2$ (middle) and $r_L = 0.0$ (right). The process $Y(\ss)$ is a Gaussian process (first row) and Student-$t$ process with 4 degrees of freedom and with Fr\'echet marginals with $\beta=1.2$ (second row), with the correlation function $\rho_Y(h) = \exp(-h/2)$.}}
		\label{Fig2}
	\end{center}
\end{figure}

Figure~\ref{Fig2} shows realizations of the process $\tilde Z(\ss)$ with the same tail dependence function $\lambda_U(h)$ as the process $Z(\ss)$ whose realizations are shown in Figure~\ref{Fig1} but with stronger overall bulk dependence as measured by $S_{\rho}(h)$. We consider two choices for $Y(\ss)$: a standard Gaussian process with $N(0,1)$ marginals (which can be considered as a limiting case with $\beta \to \infty$) and the Student-$t$ process with Fr\'echet marginals $F_Y(z) = \exp(-z^{-\beta})$, $\beta=1.2$. We can see that realizations of $\tilde Z(\ss)$ for these two choices of $Y(\ss)$ look relatively similar. However, in the first row, the new process has tail quadrant independence at large distances, i.e., $\tilde\kappa_U=2$, while in the second row, the process has intermediate tail independence with tail order $\tilde\kappa_U = 1.2$.  

We now derive the lower tail dependence structure of the process $\tilde Z(\ss)$.  Assuming $C_{\YY}$ has lower tail-dependence with $\lambda_L^{\YY} > 0$, we can write:
$$
\Pr(\tilde Z(\ss_1) < z, \tilde Z(\ss_2) < z) \sim K_Le^{-q\kappa_L^{\ZZ}/z}F_Y\left(\frac{z}{1-q}\right)\,, \quad \Pr(\tilde Z(\ss_1) < z) =  e^{-q/z}F_Y\left(\frac{z}{1-q}\right), \quad z \downarrow 0,
$$
where $\kappa_L^{\ZZ} = 2 - \delta_1(r_U, r_U)$ is the lower tail order of the copula $C_{\ZZ}$ defined in Section~\ref{subsec1.2} and $K_L > 0$ is some constant. It follows that $\tilde C$ is tail-independent in its lower tail. Furthermore, if $F_Y(0) > 0$ (e.g., $Y$ follows a Student's-$t$ distribution), then  $\tilde\kappa_L = \kappa_L^{\ZZ}$.  Similar results hold if $\lambda_L^{\YY} = 0$. On the other hand, if $F_Y(0) = 0$, $\tilde\kappa_L$ depends on the behavior of $F_Y(z)$ around zero. In particular, if $F_Y(z)$ converges to zero at a faster rate then $e^{-1/z}$, e.g., if $F_Y(z) = \exp(-z^{-\beta})$, $z > 0$, $\beta > 1$, then it is easy to see that $\tilde\kappa_L = \kappa_L^{\YY}$. One interesting special case arises when $C_{\YY}$ has lower tail-dependence, which implies $\tilde\kappa_L = 1$, so that the copula $\tilde C$ has quite strong dependence in the lower tail in this boundary case.

\section{Inference}
\label{sec-infer}
We now discuss inference methods for the max-convolution process $Z(\ss)$ as defined in Section~\ref{subsec1.2} and the max-mixture process $\tilde Z(\ss)$ from Section~\ref{subsec1.3} that provides greater flexibility in the bulk of the joint distribution as well as its tails. 

\subsection{Max-convolution process in Section~\ref{subsec1.2}}
\label{subsec-infer1}
Consider a sample $\{(z_{i1},\ldots,z_{ip})^{\top}\}_{i=1}^n$ where $(z_{i1},\ldots,z_{ip})^{\top}$ are i.i.d. realizations of $(Z(\ss_1), \ldots, Z(\ss_p))^{\top}$ from the process $Z(\ss)$ in (\ref{eq2s}), for $i=1,\ldots, n$. Note that the copula linking  $(Z(\ss_1), \ldots, Z(\ss_p))^{\top}$ can be combined with arbitrary univariate marginals, which, in the general case, are unknown. If the parametric form of marginal distributions is known, then their parameters can be estimated using the marginal likelihood approach and then the integral transform can be applied to get the same data but with $U(0,1)$ marginals, which we denote by $\{(u_{i1},\ldots,u_{ip})^{\top}\}_{i=1}^n$. Alternatively, a nonparametric approach based on ranks can also be used \citep{Genest.Ghoudi.ea1995}. The copula parameters can be estimated in the second step; such a two-step estimation approach is computationally fast and yields consistent and asymptotically normal estimates of the marginal and copula parameters under mild conditions \citep{Joe.Xu1996, Joe2005}.  

The full joint copula density of $(Z(\ss_1), \ldots, Z(\ss_p))^{\top}$ is difficult to compute. To circumvent this problem, the composite likelihood approach can be used to estimate copula parameters \citep{Lindsay.1998,Varin.Reid.ea2011,Varin.Vidoni2005}. Let $C_{\bm Z}^{j_1,j_2}(\cdot,\cdot)$ be the copula cdf that links $Z(\ss_{j_1})$ and $Z(\ss_{j_2})$. If $r_L \neq r_U$, the respective copula density $c_{\bm Z}^{j_1,j_2}(\cdot,\cdot)$ exists, and we can thus define a pairwise log-likelihood as
\begin{equation}
\label{eq-complik}
\ell_p(\uu; \tht) = \sum_{i=1}^n\sum_{j_1 < j_2} w_{j_1,j_2}\ln c_{\bm Z}^{j_1,j_2}(u_{ij_1}, u_{ij_2}; \tht),
\end{equation}
where $w_{j_1,j_2} \geq 0$ are some weights, and $\tht$ is a vector of unknown parameters that includes $r_L$, $r_U$ as well as the parameters of the pdf $g_R$ that links $R(\ss_{j_1})$ and $R(\ss_{j_2})$. The pairwise likelihood estimator $\widehat\tht = \mathrm{argmax}_{\tht} \,\ell_p(\uu; \tht)$ is consistent and asymptotically normal under standard regularity conditions \citep{Lindsay.1998, Varin.Reid.ea2011}. 

Note that one cannot differentiate under the integration sign in (\ref{eq-copcdf}) to compute the density $c_{\bm Z}^{j_1,j_2}(u_1,u_2)$, so one possibility is instead to estimate the density numerically as
\begin{eqnarray*}
c_{\bm Z}^{j_1,j_2}(u_1,u_2) &\approx  &\frac{1}{2\epsilon}\left\{C_{\bm Z}^{j_1,j_2}(u_1+\epsilon,u_2+\epsilon) + C_{\bm Z}^{j_1,j_2}(u_1-\epsilon,u_2-\epsilon)\right\}\\
  &&-\frac{1}{2\epsilon}\left\{C_{\bm Z}^{j_1,j_2}(u_1+\epsilon,u_2-\epsilon) - C_{\bm Z}^{j_1,j_2}(u_1-\epsilon,u_2+\epsilon)\right\}\,,
\end{eqnarray*}
where $\epsilon > 0$ is a small positive real, and the copula cdf can be calculated using numerical integration, e.g., using Gauss--Legendre quadrature \citep{Stroud.Secrest1966}. However, this approach requires $n_q > 100$ quadrature points to produce accurate results, so the computation can be very slow.  

Instead, the copula density can be re-written as a two-dimensional integral, similar to the copula cdf, and it only requires $n_q = 35$ quadrature points to compute the integral with a good accuracy; more details are provided in the Appendix.

\subsection{Max-mixture process in Section~\ref{subsec1.3}}
\label{subsec-infer2}
Again, consider a sample $\{(\tilde z_{i1},\ldots,\tilde z_{ip})^{\top}\}_{i=1}^n$ where $(\tilde z_{i1},\ldots,\tilde z_{ip})^{\top}$ are i.i.d. realizations of $(\tilde Z(\ss_1), \ldots, $ $\tilde Z(\ss_p))^{\top}$ for $i=1,\ldots, n$, from the process defined in (\ref{eq-convproc}).  

Let $\tilde C^{j_1,j_2}(\cdot,\cdot)$ be the copula cdf linking $\tilde Z(\ss_{j_1})$ and $\tilde Z(\ss_{j_2})$ and $\tilde c^{j_1,j_2}(\cdot,\cdot)$ be the respective copula pdf. Similar to the original process, the data can be marginally transformed to the $U(0,1)$ scale, and we denote these data $\{(u_{i1}, \ldots, u_{ip})^{\top}\}_{i=1}^n$; a similar pairwise log-likelihood function can be used to estimate the model parameters, and it can be written as
\begin{equation}
\label{eq-Gausscomplik}
\tilde\ell_p(\uu; \tilde\tht) = \sum_{i=1}^n\sum_{j_1 < j_2} w_{j_1,j_2}\ln \tilde c^{j_1,j_2}(u_{ij_1}, u_{ij_2}; \tilde \tht),
\end{equation}
where $w_{j_1,j_2} \geq 0$ are some weights, and $\tilde\tht$ is a vector of unknown parameters that includes the same parameters as for the original max-convolution process $Z(\ss)$, as well as the parameter $q$ and parameters controlling the process $Y(\ss)$. Again, the composite likelihood estimator $\widehat{\tilde\tht} = \mathrm{argmax}_{\tilde\tht} \,\tilde\ell_p(\uu; \tilde\tht)$ is consistent and asymptotically normal under standard regularity conditions. 
 
Let $\tilde F^{j_1,j_2}$ and $\tilde f^{j_1,j_2}$ be the joint cdf and pdf, respectively, of $(\tilde Z(\ss_{j_1}), \tilde Z(\ss_{j_2}))^{\top}$. Note that 
\begin{eqnarray*}
\tilde F^{j_1,j_2}(z_1, z_2) &=& \Pr\left\{Z(\ss_{j_1}) \leq \frac{z_1}{q}, Z(\ss_{j_2}) \leq \frac{z_2}{q}\right\}\cdot \Pr\left\{Y(\ss_{j_1}) \leq \frac{z_1}{1-q}, Y(\ss_{j_2}) \leq \frac{z_2}{1-q}\right\}\\
&=& C_{\bm Z}^{j_1,j_2}\left\{\exp(-q/z_1), \exp(-q/z_2)\right\}\cdot F_{\bm Y}^{j_1,j_2}\left(\frac{z_1}{1-q}, \frac{z_2}{1-q}\right),
\end{eqnarray*}
where $C_{\bm Z}^{j_1,j_2}$ is the copula linking $(Z(\ss_{j_1}), Z(\ss_{j_2}))^{\top}$, and $F_{\bm Y}^{j_1,j_2}$ is the joint cdf of $(Y(\ss_{j_1}), Y(\ss_{j_2}))^{\top}$. 
This implies that 
$$
\tilde C^{j_1,j_2}(u_1, u_2) = \tilde F^{j_1,j_2}\left\{\tilde F^{-1}(u_1), \tilde F^{-1}(u_2)\right\}, \quad \tilde c^{j_1,j_2}(u_1, u_2) = \frac{\tilde f^{j_1, j_2}\left\{\tilde F^{-1}(u_1), \tilde F^{-1}(u_2)\right\}}{\tilde f\left\{\tilde F^{-1}(u_1)\right\} \tilde f\left\{\tilde F^{-1}(u_2)\right\}}\,,
$$
where $\tilde F(z) = e^{-q/z}\cdot F_Y\left\{z/(1-q)\right\}$, $\tilde F^{-1}$ is the inverse marginal cdf, and where the marginal pdf is
$$\tilde f(z) = \tilde F'(z) = \frac{q}{z^2}e^{-q/z}\cdot F_Y\left(\frac{z}{1-q}\right) + \frac{1}{1-q}e^{-q/z}\cdot f_Y\left(\frac{z}{1-q}\right), \quad f_Y(z) = F_Y'(z).$$
The inverse cdf $\tilde F^{-1}(\cdot)$ can be easily computed using numerical methods, e.g., a bisection method.

\section{Simulation studies}
\label{sec-sim}

In this section, we assess the performance of the pairwise likelihood  estimators proposed in Section~\ref{sec-infer}. We use $p=10, 20, 30, 50$ randomly selected locations in $[0,1]^2$ for each simulated data set. For each simulation study, we simulate $N=200$ data sets with $n=100$ and $n=500$ independent replicates.

\subsection{Simulation study 1}
\label{subsec-sim1}

We first consider the process $Z(\ss)$ as defined in (\ref{eq2s}) in Section~\ref{subsec1.2}.  We use a Gaussian process $R(\ss)$ defined in $\mathbb{R}^2$ with uniform $U(0, r)$ marginals, which corresponds to a process with quite rough sample paths, and exponential covariance function $\rho_R(h;\theta_R) = \exp(-h/\theta_R)$. We select $\tht = (r, \theta_R)^{\top} = (0.4, 0.25)^{\top}$, but similar results can be obtained with different parameter values. 

We use the pairwise likelihood method as explained in Section~\ref{subsec-infer1} and we use Gauss--Legendre quadrature with $n_q = 35$ quadrature points to compute the bivariate copula pdf in (\ref{eq-complik}). To make computations faster and to remove pairs with very weak dependence, we set $w_{j_1,j_2} = 1$ if $||\ss_{j_1} - \ss_{j_2}|| < 0.25$ and $w_{j_1,j_2} = 0$ otherwise. We assume here that univariate marginals are unknown and we use the nonparametric approach based on ranks to transform data to the uniform scale. Table~\ref{tab-sim1} reports the results.
\begin{table}[t!]
    \caption{{Simulation study 1: Root Mean Square Errors (RMSEs) of estimates of the copula parameters $\tht = (r, \theta_R)^{\top}$ obtained using the  pairwise likelihood estimator considered in Section~\ref{sec-infer}. True parameters are set to $(r, \theta_R)^{\top} = (0.4, 0.25)^{\top}$. The results are based on $N=200$ simulated data sets with $n = 100$ and $n=500$ replicates and $p=10, 20, 30, 50$ randomly selected locations in $[0,1]^2$.}}
	\label{tab-sim1}
	\def~{\hphantom{0}}
	\begin{center}
		{\begin{tabular}{l|c|c|c|c}
				& $p=10$ & $p=20$ & $p=30$ & $p=50$\\
				\hline
				$n = 100$ & (0.13, 0.70) & (0.05, 0.08) & (0.04, 0.06) & (0.04, 0.05)\\
				$n = 500$ & (0.05, 0.06) & (0.02, 0.03) & (0.02, 0.03) & (0.02, 0.02)\\		
		\end{tabular}}		
	\end{center}
\end{table}

As expected, the pairwise likelihood estimates improve as $p$ and $n$ increase. In particular, they have much smaller RMSEs, especially for the covariance function parameter $\theta_R$, when $p \geq 20$. 

\subsection{Simulation study 2}
\label{subsec-sim2}

We now consider the process $\tilde Z(\ss)$ as defined in (\ref{eq-convproc}) in Section~\ref{subsec1.3}. We use the same parameters for the process $Z(\ss)$ as in Section~\ref{subsec-sim1}, and use a Gaussian process $Y(\ss)$ with $N(0,1)$ marginals (which can be considered as a special case with $\beta \to \infty$) and exponential covariance function $\rho_Y(h; \theta_Y) = \exp(-h/\theta_Y)$ with $\theta_Y = 0.5$, and $q = 0.2$.

We use the pairwise likelihood approach as explained in Section~\ref{subsec-infer2} and we employ Gauss--Legendre quadrature with $n_q = 35$ quadrature points to compute the bivariate copula pdf in (\ref{eq-Gausscomplik}). Similar to the first simulation in Section~\ref{subsec-sim1}, we set $w_{j_1,j_2} = 1$ if $||\ss_{j_1} - \ss_{j_2}|| < 0.25$ and $w_{j_1,j_2} = 0$ otherwise. We again assume that univariate marginals are unknown and we use the nonparametric approach based on ranks to transform data to the uniform scale. Table~\ref{tab-sim2} reports the results.

\begin{table}[t!]
    \caption{{Simulation study 2: Root Mean Square Errors (RMSEs) of the estimates of copula $\tilde\tht = (r, \theta_R, \theta_Y, q)^{\top}$ obtained using the  pairwise likelihood estimator considered in Section~\ref{sec-infer}. The true values are set to $\tilde\tht = (0.4, 0.25, 0.5, 0.2)^{\top}$. The results are based on $N=200$ simulated data sets with $n = 100$ and $n=500$ replicates and $p=10, 20, 30, 50$ randomly selected locations in $[0,1]^2$.}}
	\label{tab-sim2}
	\def~{\hphantom{0}}
	\begin{center}
		{\begin{tabular}{l|c|c}
				& $p=10$ & $p=20$\\ 	
				\hline
				$n = 100$ & (0.25, 1.55, 0.26, 0.06) & (0.08, 0.15, 0.13, 0.04) \\
				$n = 500$ & (0.11, 0.08, 0.08, 0.02) & (0.04, 0.05, 0.04, 0.01) \\
                \hline
    		& $p=30$ & $p=50$\\ 	
                \hline
                $n=100$ &(0.07, 0.11, 0.09, 0.03) & (0.05, 0.10, 0.09, 0.03)\\
                $n=500$ &(0.03, 0.04, 0.04, 0.01) & (0.03, 0.04, 0.04, 0.01)	
	    \end{tabular}}		
	\end{center}
\end{table}

Again, we can see that RMSEs are much smaller if $p \geq 20$, and estimates are more accurate if a larger sample size is used. In both cases, the estimates are quite accurate, even when $n$ is rather small provided $p \geq 20$.

\subsection{Simulation study 3}
\label{subsec-sim3}

Finally, we again consider the process $\tilde Z(\ss)$ as defined in (\ref{eq-convproc}) in Section~\ref{subsec1.3}, but in a more challenging scenario. We use the same parameters as for the process $Z(\ss)$ in Section~\ref{subsec-sim1}, but now taking a Student's-$t$ process for $Y(\ss)$, with $\nu = 3$ degrees of freedom and Fr\'echet marginals $F_Y(z) = \exp(-z^{-\beta})$, $z > 0$ with $\beta = 1.2$ and exponential covariance function $\rho_Y(h; \theta_Y) = \exp(-h/\theta_Y)$, with $\theta_Y = 0.5$, and $q = 0.2$.

This time, the parameter $\beta = 1.2$ controls the strength of dependence at larger distances and therefore only selecting pairs at shorter distances as in the previous two simulation studies may result in poor estimates for this parameter. Indeed, using the pairwise likelihood with $w_{j_1, j_2} = 1$ if $||\ss_{j_1} - \ss_{j_2}|| < 0.25$ and $w_{j_1,j_2} = 0$ otherwise yields very poor estimates of the model parameters. Much better results are achieved when $w_{j_1, j_2} = 1$ if $||\ss_{j_1} - \ss_{j_2}|| <  0.5$, and with $w_{j_1,j_2} = 0$ otherwise. We again use the nonparametric approach based on ranks to transform the data, and we do not estimate the $\nu$ parameter, which is difficult to infer together with the other parameters. We do not restrict the parameter $\beta$ to be greater than one and let the data decide whether $\beta \leq 1$ or $\beta > 1$, which corresponds to  long-range and short-range upper tail-dependence, respectively. Table~\ref{tab-sim3} reports the results.

\begin{table}[t!]
    \caption{{Simulation study 3: Root Mean Square Errors (RMSEs) of the estimates of copula $\tilde\tht = (r, \theta_R, \theta_Y, \beta, q)^{\top}$ obtained using the  pairwise likelihood estimator considered in Section~\ref{sec-infer} and weights $w_{j_1,j_2} = 1$ if $||\ss_{j_1} - \ss_{j_2}|| < 0.5$ and $w_{j_1,j_2} = 0$ otherwise. The true values are set to $\tilde\tht = (0.4, 0.25, 0.5, 1.2,  0.2)^{\top}$. The results are based on $N=200$ simulated data sets with $n = 100$ and $n=500$ replicates and $p=10, 20, 30, 50$ randomly selected locations in $[0,1]^2$.}}
	\label{tab-sim3}
	\def~{\hphantom{0}}
	\begin{center}
		{\begin{tabular}{l|c|c}
				 & $p=10$ & $p=20$ \\ 	
				\hline
				$N = 100$ & (0.90, 3.37, 0.13, 0.61, 0.10) & (1.06, 3.66, 0.14, 0.38, 0.05) \\
				$N = 500$ & (0.35, 2.82, 0.07, 0.16, 0.05) & (0.30, 2.85, 0.06, 0.12, 0.04) \\		
                \hline
                & $p=30$ & $p=50$ \\
                \hline 
                $N = 100$ &  (1.00, 3.94, 0.15, 0.59, 0.06) & (1.08, 3.81, 0.15, 0.33, 0.06)\\
				$N = 500$ & (0.23, 2.79, 0.06, 0.10, 0.04) & (0.32, 3.24, 0.08, 0.12, 0.05) \\		
	    \end{tabular}}		
	\end{center}
\end{table}

The results are less accurate for this model, which indicates that parameters are weakly identifiable in this setting. The value $\beta = 1.2$ is quite close to one, so the two mixture components in (\ref{eq-convproc}) play a quite similar role, and a larger sample size is required to obtain reasonable parameter estimates. Parameter estimates are more accurate with a larger sample size, though the improvement is small for $\theta_R$. Using more spatial locations helps to somewhat improve estimates of $\beta$ and $q$, but does not seem to affect the results for the remaining parameters. One reason is that more pairs at longer distances are now selected, so adding more locations does not significantly improve parameter estimates. Furthermore, all the parameters $(r, \theta_R, \theta_Y, \beta, q)^{\top}$ affect the tail properties of the process considered in this section, unlike the process $\tilde Z(\ss)$ used in Section \ref{subsec-sim2}, where only the first two parameters $r$ and $\theta_R$ determine the tail behavior, and the remaining parameters $\theta_Y$ and $q$ affect the behavior in the bulk of the joint distribution. As a result, it is easier to identify all the parameters in the latter case.

Although the parameter estimates are less accurate in this setting, the dependence structure is estimated very well. For example, one of the estimated values of $\tilde\tht = (r, \theta_R, \theta_Y, \beta, q)^{\top}$ we obtained for a simulated data set is $\widehat\tht = (0.39, 5, 0.42, 0.87, 0.125)^{\top}$. To assess the fit of the estimated model in the joint lower and upper tails, we use tail-weighted measures of dependence proposed by \cite{Krupskii.Joe2015} (denoted by $\varrho_L$ and $\varrho_U$, respectively) for each pair of variables. In addition, we use the Spearman's rho (denoted by $S_{\rho}$) for each pair of variables. We compute the absolute differences between these quantities for the true model with $\tilde\tht = (0.4, 0.25, 0.5, 1.2, 0.2)^{\top}$ and the estimated model with $\widehat\tht = (0.39, 5, 0.42, 0.87, 0.125)^{\top}$, averaged across different pairs of variables, denoted $|\Delta|_{S_{\rho}}$, $|\Delta|_{\varrho_L}$ and $|\Delta|_{\varrho_U}$, respectively.  We find that $|\Delta|_{S_{\rho}} = 0.02$, $|\Delta|_{\varrho_L} = 0.03$ and $|\Delta|_{\varrho_U} = 0.04$ which indicates a very accurate fit both in the tails and in the bulk of the distribution.

\section{Temperature data application}
\label{sec-empstudy}

We apply the proposed methodology to analyze temperature data measured at $p=100$ stations in the state of Oklahoma, United States. We use daily maxima, and the time period is May 1, 2022 to September 30, 2022, which contains $n=153$ days in total. We do not include winter data as the weather patterns can be considerably different during winter and summer months. The data can be downloaded from the website \texttt{mesonet.org}. We remove the seasonal component and fit an AR(2) model to remove the temporal dependence. We then transform the residuals to the uniform $U(0,1)$ scale using nonparametric ranks. 

\begin{figure}[t!]
    \centering
    \includegraphics[width=12cm]{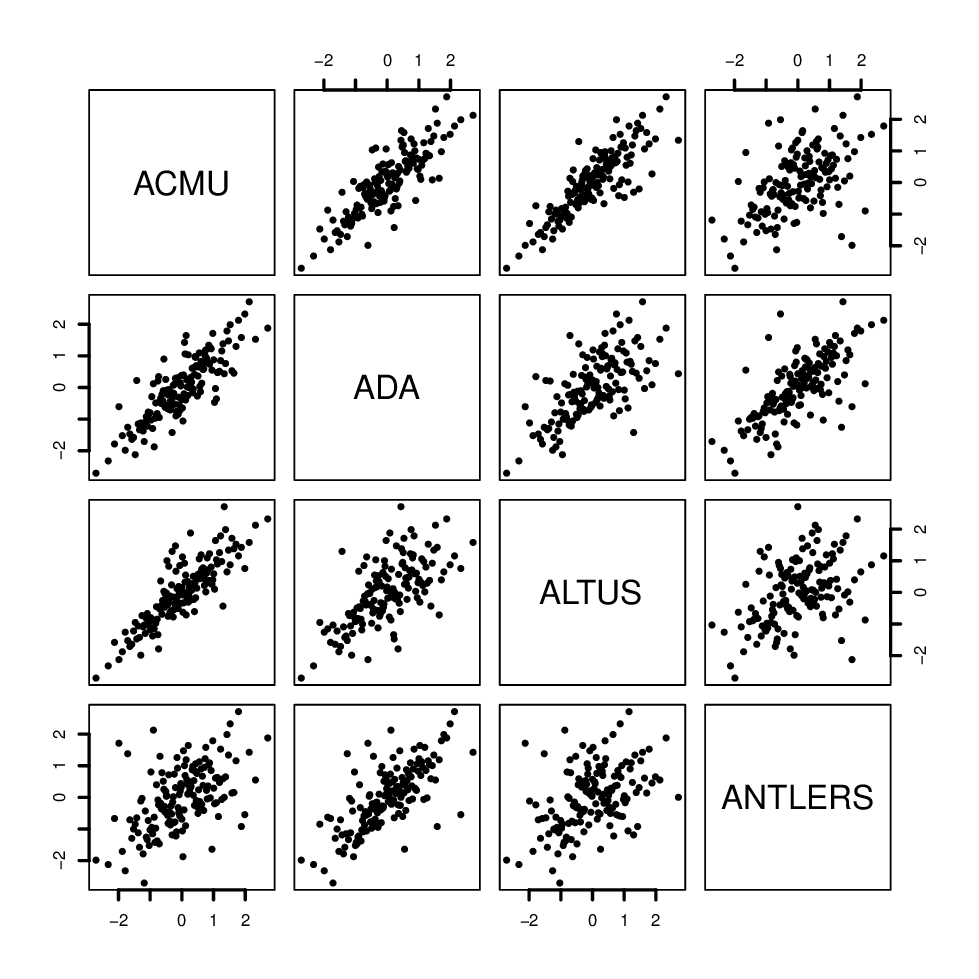}
    \caption{Normal scores scatter plots of the AR(2) residuals obtained using the daily maxima of the temperatures recorded between May 1, 2022 and September 30, 2022 in Oklahoma, US, for some selected pairs of stations.}
    \label{fig1}
\end{figure}

Figure~\ref{fig1} shows scatter plots of residuals transformed to the standard normal $N(0,1)$ marginals for some selected pairs of stations. Asymmetric dependence can be observed in the scatter plots, with a stronger dependence in the joint lower tail. 
To confirm these findings, we use tail-weighted measures of dependence $\varrho_L$ and $\varrho_U$ we used in the previous section to assess the strength of dependence in the joint lower and upper tails for each pair of stations. We also compute the parametric estimates of these measures under the assumption of a Gaussian copula (denoted by $\varrho_N$ as the value of this measure is the same in the lower and upper tail for this copula).
In addition, we compute the Spearman's rho (denoted by $S_{\rho}$) for each pair of stations. Table~\ref{tab:data_summary} shows the values of $S_{\rho}$, $\varrho_L$, $\varrho_U$, and $\varrho_N$, averaged across different pairs of stations at distances $h < 100$, $100 \leq h < 200$, $200 \leq h < 300$, $300 \leq h < 400$, $400 \leq h < 500$ and $h \geq 500$, where $h$ is measured in kilometers. The strength of dependence, as measured by $S_{\rho}$, $\varrho_L$ and $\varrho_U$ gets weaker with larger distances, as expected; however the dependence in the joint lower tail is stronger than that of the normal copula, especially at larger distances, whereas the dependence in the joint upper tail is close to that of the normal copula. It implies that models based on symmetric copulas, such as the normal or Student's-$t$ copula may not be suitable for these data.

\begin{table}[h]
    \centering
    \caption{Measures $S_{\rho}, \varrho_L, \varrho_U$ and $\varrho_N$ calculated for the daily maximum temperatures, averaged across different pairs of stations at distance $h$ from each other (measured in km), for different values of $h$}
    \label{tab:data_summary}
    \begin{tabular}{c|c|c|c|c|c|c}
    \multirow{2}{*}{measure} & \multicolumn{6}{c}{distance $h$}\\
    \cline{2-7}
    & $< 100$& $(100, 200)$ & $(200, 300)$ & $(300, 400)$ & $(400, 500)$ & $\geq 500$\\
    \hline
     $S_{\rho} $    &0.90&0.75&0.61&0.49&0.39&0.26\\
     $\varrho_L$    &0.89&0.74&0.59&0.51&0.43&0.34\\
     $\varrho_U$    &0.79&0.57&0.39&0.30&0.23&0.17\\
     $\varrho_N$    &0.78&0.54&0.37&0.26&0.19&0.11
    \end{tabular}
\end{table}

We now fit several copula models to these data:

\begin{itemize}
    \item[M1:] Gaussian copula with a powered-exponential covariance function $\rho(h; \theta, \alpha) = \exp\left\{-(h/\theta)^{\alpha}\right\}$, $\theta > 0$, $\alpha \in (0, 2]$;
    \item[M2:] Student's-$t$ copula with a powered exponential covariance function $\rho(h; \theta, \alpha) = \exp\left\{-(h/\theta)^{\alpha}\right\}$, $\theta > 0$, $\alpha \in (0, 2]$, and $\nu > 0$ degrees of freedom;
    \item[M3:] The copula corresponding to the process $Z(\ss)$ as defined in (\ref{eq2s}) with the Gaussian process $R(\ss)$ with uniform $U(0,r)$ marginals, $r > 0$, and exponential covariance function $\rho_R(h; \theta_R) = \exp(-h/\theta_R)$, $\theta_R > 0$;
    \item[M4:] The copula corresponding to the process $\tilde Z(\ss)$ as defined in (\ref{eq-convproc}), with the Gaussian process $R(\ss)$ with uniform $U(0,r)$ marginals, $r > 0$, and exponential covariance function $\rho_R(h; \theta_R) = \exp(-h/\theta_R)$, $\theta_R > 0$, and a Gaussian process $Y(\ss)$ with standard normal marginals and powered exponential covariance function $\rho_Y(h; \theta_Y; \alpha_Y) = \exp\left\{-(h/\theta_Y)^{\alpha_Y}\right\}$, $\theta_Y > 0, \alpha_Y \in (0,2]$;
    \item[M5:] The copula corresponding to the process $\tilde Z(\ss)$ as defined in (\ref{eq-convproc}), with the Gaussian process $R(\ss)$ with uniform $U(0,r)$ marginals, $r > 0$, and exponential covariance function $\rho_R(h; \theta_R) = \exp(-h/\theta_R)$, $\theta_R > 0$, and a Student's-$t$ process $Y(\ss)$ with $\nu = 4$ degrees of freedom and Fr\'echet marginals $F_Y(z) = \exp\left(-z^{-\beta}\right)$, $\beta > 0$, and powered exponential covariance function $\rho_Y(h; \theta_Y; \alpha_Y) = \exp\left\{-(h/\theta_Y)^{\alpha_Y}\right\}$, $\theta_Y > 0, \alpha_Y \in (0,2]$.
\end{itemize}

Since the data show stronger dependence in the joint lower tail, and models M3--M5 can capture stronger dependence in the upper tail, we fit these models to the negated residuals. Since the parameters in model M5 are weakly identifiable as we showed in Section \ref{subsec-sim3}, we fix $\theta_R = 5$ for this model.

To assess the out-of-sample performance of these models, we randomly select between 70 and 90 stations to estimate parameters of these models. For the estimated models, we use Monte Carlo simulations to approximate the values of $S_{\rho}, \varrho_L$ and $\varrho_U$ for each pair of the remaining stations. We compute the difference between the empirical and model-based estimates of these measures for the remaining stations, averaged across different pairs of variables at distances $h < 100$, $100 \leq h < 200$, $200 \leq h < 300$, $300 \leq h < 400$, $400 \leq h < 500$ and $h \geq 500$, where $h$ is measured in kilometers. We repeat this procedure 10 times; and Table~\ref{tab:data_fit} shows the results averaged across 10 repetitions.

\begin{table}[h]
    \centering
    \caption{Out-of-sample differences between the empirical and model-based estimates of $S_{\rho}$, $\varrho_L$, and $\varrho_U$, averaged across different pairs of variables at distances $h < 100$, $100 \leq h < 200$, $200 \leq h < 300$, $300 \leq h < 400$, $400 \leq h < 500$ and $h \geq 500$, computed for models M1--M5. Parameter estimates are calculated using randomly selected 70--90 stations, and the differences are computed using the remaining stations; this procedure is repeated 10 times, and the results are averaged across 10 repetitions. Here $h$ is measured in kilometers. Note that copulas are here fitted to the negated residuals, so the lower tail corresponds to the upper tail of the original data, and vice versa.}
    \label{tab:data_fit}
    \begin{tabular}{ccccccc}
    \multirow{2}{*}{} & \multicolumn{6}{c}{distance $h$}\\
    \cline{2-7}
    & $< 100$& $(100, 200)$ & $(200, 300)$ & $(300, 400)$ & $(400, 500)$ & $\geq 500$\\
    \cline{2-7}
    Model &\multicolumn{6}{c}{Spearman's rho, $S_{\rho}$}\\
    \hline
    M1&$-0.01$&$-0.03$&$-0.05$&$-0.07$&$-0.07$&$-0.08$\\
    M2&$-0.02$&$-0.05$&$-0.09$&$-0.12$&$-0.12$&$-0.13$\\
    M3&~~0.13&~~0.08&~~0.00&$-0.08$&$-0.11$&$-0.16$\\
    M4&~~0.07&~~0.08&~~0.09&~~0.08&~~0.08&~~0.04\\
    M5&~~0.03&~~0.03&~~0.03&~~0.02&~~0.02&$-0.02$\\
    \hline
    &\multicolumn{6}{c}{Measure of dependence in the lower tail, $\varrho_{L}$}\\
    \hline
    M1&~~0.10&~~0.15&~~0.16&~~0.18&~~0.16&~~0.10\\
    M2&~~0.06&~~0.07&~~0.05&~~0.05&~~0.04&~~0.00\\
    M3&~~0.10&~~0.05&$-0.04$&$-0.07$&$-0.10$&$-0.12$\\
    M4&~~0.10&~~0.13&~~0.15&~~0.16&~~0.14&~~0.08\\
    M5&~~0.03&~~0.04&~~0.02&~~0.03&~~0.02&$-0.01$\\
    \hline
    &\multicolumn{6}{c}{Measure of dependence in the upper tail, $\varrho_{U}$}\\
    \hline
    M1&~~0.00&$-0.01$&$-0.04$&$-0.01$&$-0.01$&$-0.03$\\
    M2&$-0.05$&$-0.09$&$-0.16$&$-0.14$&$-0.14$&$-0.13$\\
    M3&~~0.01&$-0.10$&$-0.22$&$-0.25$&$-0.26$&$-0.24$\\
    M4&~~0.22&~~0.13&~~0.02&$-0.01$&$-0.05$&$-0.04$\\
    M5&~~0.03&~~0.02&$-0.03$&$-0.01$&$-0.02$&$-0.04$\\
    \hline
    \end{tabular}
\end{table}

Model~M1 tends to overestimate dependence in the bulk of distribution, as measured by $S_{\rho}$, at longer distances, and underestimate it in the lower tail as expected. Model~M2 tends to overestimate the overall dependence as well as dependence in the upper tail. Model~M3 has quite a rigid dependence structure so it is not flexible enough to accurately model dependence observed in the data, especially at larger distances for the upper tail. On the other hand, Model~M4 tends to slightly underestimate dependence in the lower tail and in the upper tail at shorter distances. Model~M5 has the best fit to the data, both in the bulk of the joint distribution, and its tails, and it can capture dependencies very well both at short distances and long distances, unlike the other four models.

\section*{Conclusion}

We have introduced a new class of models for spatial data and showed that these models can handle data with complex dependence structures, including tail-dependence at short distances and tail-independence at larger distances, with exact independence at infinite distances. Furthermore, the full range of dependence can be achieved at long distances, from tail-dependence to intermediate tail-dependence or tail-quadrant-independence. This class of models can capture tail asymmetry, and the model parameters can be computed using a weighted pairwise likelihood approach. We have shown in simulation studies that accurate parameter estimates can be obtained in most cases, provided the number of replicates and stations is large enough.

While the model parameters can be efficiently estimated using the weighted pairwise likelihood approach, the joint copula density of the process is not tractable in the general case and it would be interesting to explore alternative likelihood-free estimation approaches, such as neural Bayes estimators \citep{Sainsbury-Dale.ea2023,Sainsbury-Dale.ea2023b,Richards.ea2023}. Moreover, simulation of the spatial process conditional on the observed values of this process at some locations
is not feasible. One could use some version of rejection sampling for conditional sampling given that simulations from the proposed model can be performed very fast.  Since conditioning on multiple values is not computationally feasible, one can use the value of a single aggregation functional. To further enhance conditional simulation, one can adapt exponential tilting and importance sampling methods \citep{Rached.etal:2016, Botev.LEcuyer:2017} especially when the conditioning event
is a low-probability rare event. 

One limitation of the proposed class of models is that they cannot capture lower tail dependence, so extensions of these models (possible involving a combination of max- and min-convolution processes) that would allow both for lower and upper tail dependence is a topic of future research.

\section*{Appendix}

We now show how to compute the density of a copula $C_{\bm Z}(u_1,u_2;h)$ in (\ref{eq-copcdf}). To illustrate the ideas, we assume that $r_L = 0$ and $r_U =1$ for simplicity. Without loss of generality, we also assume $u_1 > u_2$.
Let $\tilde u_i = - \ln u_i$, $i=1,2$, and $\ell_{12} = (\tilde u_1/ \tilde u_2)^{1/2}$.
Note that
$$C_{\bm Z}(u_1,u_2;h) = \int_{u_1^{\delta_1} < u_2^{\delta_2}} (u_1 u_2^{1-\delta_2} - u_2 u_1^{1-\delta_1}) g_W(w_1, w_2; h) \d w_1 \d w_2 + \int_{[0,1]^2} u_2 u_1^{1-\delta_1} g_W(w_1, w_2; h) \d w_1 \d w_2,$$
where the first integral in the right hand side is
\begin{eqnarray*}
	I_0(u_1,u_2;h)	&=& \int_{w_1 < w_2 \ell_{12}} (u_1 u_2^{1-\delta_2} - u_2 u_1^{1-\delta_1}) g_W(w_1, w_2; h) \d w_1 \d w_2\\
	&= & \int_{[0,1]^2}  (u_1 u_2^{1-\delta_2(w_1w_2\ell_{12}, w_2; h)} - u_2 u_1^{1-\delta_1(w_1w_2\ell_{12},w_2;h)}) g_W(w_1w_2\ell_{12}, w_2; h) \d w_1 \d w_2.
\end{eqnarray*} 
Note that $(u_1u_2^{1-\delta_2} - u_2u_1^{1-\delta_1}) = 0$ if $w_1 = w_2\ell_{12}$. This implies that
\begin{eqnarray*}
\frac{\p C(u_1,u_2;h)}{\p u_1} &=& \frac{\p I_0(u_1,u_2;h)}{\p u_1} + \int_{[0,1]^2} (1-\delta_1)u_2u_1^{-\delta_1} g_W(w_1, w_2; h) \d w_1 \d w_2,\\
\frac{\p C(u_1,u_2; h)}{\p u_2} &=& \frac{\p I_0(u_1,u_2;h)}{\p u_2} + \int_{[0,1]^2} u_1^{1-\delta_1} g_W(w_1, w_2; h) \d w_1 \d w_2,\\
\frac{\p^2 C(u_1,u_2;h)}{\p u_1\p u_2} &=& \frac{\p^2 I_0(u_1,u_2;h)}{\p u_1 \p u_2} + \int_{[0,1]^2} (1-\delta_1)u_1^{-\delta_1} g_W(w_1, w_2; h) \d w_1 \d w_2,\\
\end{eqnarray*}
where, with $\delta_i = \delta_i(w_1w_2\ell_{12},w_2;h)$ and $\tilde\delta_i = \delta_i(w_1\ell_{12},w_1;h)$,
\begin{eqnarray*}
\frac{\p I_0(u_1,u_2;h)}{\p u_1} &=& \int_{[0,1]^2}  \left\{u_2^{1-\delta_2} - (1-\delta_1)u_2u_1^{-\delta_1}\right\} g_W(w_1w_2\ell_{12}, w_2; h) \d w_1 \d w_2,\\
\frac{\p I_0(u_1,u_2;h)}{\p u_2} &=& \int_{[0,1]^2}  \left\{(1-\delta_2)u_1 u_2^{-\delta_2} - u_1^{1-\delta_1}\right\} g_W(w_1w_2\ell_{12}, w_2; h) \d w_1 \d w_2,\\
\frac{\p^2 I_0(u_1,u_2;h)}{\p u_1 \p u_2} &=& \int_{[0,1]^2}  \left\{(1-\delta_2) u_2^{-\delta_2} - (1-\delta_1)u_1^{-\delta_1}\right\} g_W(w_1w_2\ell_{12}, w_2; h) \d w_1 \d w_2\\
&& - \frac{\ell_{12}}{2u_1\tilde u_1}\int_0^1 \left\{(1-\tilde\delta_2)u_1 u_2^{-\tilde\delta_2} - u_1^{1-\tilde\delta_1}\right\} g_W(w_1\ell_{12}, w_1; h) \d w_1.
\end{eqnarray*}
These formulas can be used to compute the copula cdf $C_{\bm Z}(u_1, u_2; h)$, its derivatives and the copula density. Gauss--Legendre quadrature can be used, with very accurate results obtained using only $n_q = 35$ quadrature points.

\end{document}